\RequirePackage[hyphens]{url}
\documentclass[aps,pra,preprint,superscriptaddress,11point]{revtex4-1}
\usepackage{enumerate}
\usepackage{listings}
\usepackage[usenames,dvipsnames,svgnames,table]{xcolor}
\usepackage{amsmath}
\usepackage{amssymb}
\usepackage{graphicx}
\usepackage{appendix}
\usepackage{eufrak}
\usepackage{dcolumn}
\usepackage{boxedminipage}
\usepackage{verbatim}
\usepackage[colorlinks=true,citecolor=blue,linkcolor=blue]{hyperref}
\usepackage{subfigure}
\usepackage{booktabs}
\begin{document}
\setlength{\parskip}{0em}
\title{Digital quantum simulation and pseudoquantum simulation  of the $\mathbb{Z}_2$ gauge-Higgs model}
\author{Yiming Ding}
\affiliation{Department of Physics  \&  State Key Laboratory of Surface Physics,   Fudan University,\\ Shanghai 200433, China}
\author{Xiaopeng Cui}
\affiliation{Department of Physics  \&  State Key Laboratory of Surface Physics,   Fudan University,\\ Shanghai 200433, China}
\author{Yu Shi}
\thanks{yushi@fudan.edu.cn}
\affiliation{Department of Physics  \&  State Key Laboratory of Surface Physics,   Fudan University,\\ Shanghai 200433, China}
\date{\today}
\begin{abstract}
We present a  quantum algorithm for digital quantum simulation of the  $\mathbb{Z}_2$ gauge-Higgs model on a $3\times 3$ lattice, which is  based on Trotter decomposition, the quantum adiabatic algorithm and its circuit realization. Then we perform a classical  demonstration, dubbed a pseudoquantum simulation,  on a GPU simulator. We obtain  useful results on this model, which suggest  the topological properties of the deconfined phase and help to clarify the phase diagram. It is suggested that the tricitical point, where the  second-order  critical lines of deconfinement-confinement transition and of  deconfinement-Higgs transition  meet, seems to be  on the the first-order critical line of confinement-Higgs transition, at a point other than the end of this critical line.
\end{abstract}
\maketitle
\section{Introduction}
\label{sec:intro}

Lattice gauge theory is a nonperturbative approach to gauge theory, especially quantum chromodynamics~\cite{lgt,Kogut,Kogut2,
lgtbooks,lgtbooks2}.  It  is important not only in particle physics, but also in condensed matter physics and even in topological quantum computing  \cite{S0217984914300129,toric}. It is usually implemented in terms of Monte Carlo simulation, however, which lacks real-time dynamics and may suffer the well-known fermion sign problem \cite{sign,sign2,troyer,signproblem}.

Recently, it has  appeared that these issues may be resolved in quantum simulation~\cite{reviews,reviews2,reviews3,
Byrnes2006,lewenstein2,Martinez2016,
Gonz_lez_Cuadra_2017,PhysRevLett.118.070501,
Kasper,Kolco,lamm,Ercolessi,Gorg,Schweizer,
Barbiero,Mil,Yang,Davoudi,
PhysRevX.10.021041,JHEP08(2020)160,
S0217979220502926,li,zhang}. Moreover, as exemplified in the quantum simulation \cite{JHEP08(2020)160,
S0217979220502926} of the pure $\mathbb{Z}_2$ gauge theory~\cite{
Wegner,fradkin,FradkinBook,Sachdev}, for quantum simulation involving only dozens of qubits, it is very useful to make a classical demonstration on a high-performance platform, which we called a pseudoquantum simulation~\cite{JHEP08(2020)160}. It  serves not only as a benchmark for experimental quantum simulation, which facilitates the development of quantum algorithms, but also as a new numerical method for computational problems.

We now go beyond the pure gauge theory, and consider  the $\mathbb{Z}_2$  gauge-Higgs model~\cite{PhysRevD.19.3682}, where there exists  coupling between matter and gauge fields, with duality between them.   This model has been widely studied   analytically \cite{PhysRevD.19.3682, Banks:1980gs,Gliozzi, Nussinov:2004ns, Vidal:2008uy} and  numerically \cite{Jongeward:1980wx,PhysRevD.21.1006,
PhysRevB.82.085114,2012.15845}. Remarkably, this model is  equivalent to the transverse-field toric code model \cite{PhysRevB.82.085114}, which is important for topological quantum computing.

\begin{figure}[h!] 
 \center{\includegraphics[width=8cm]  {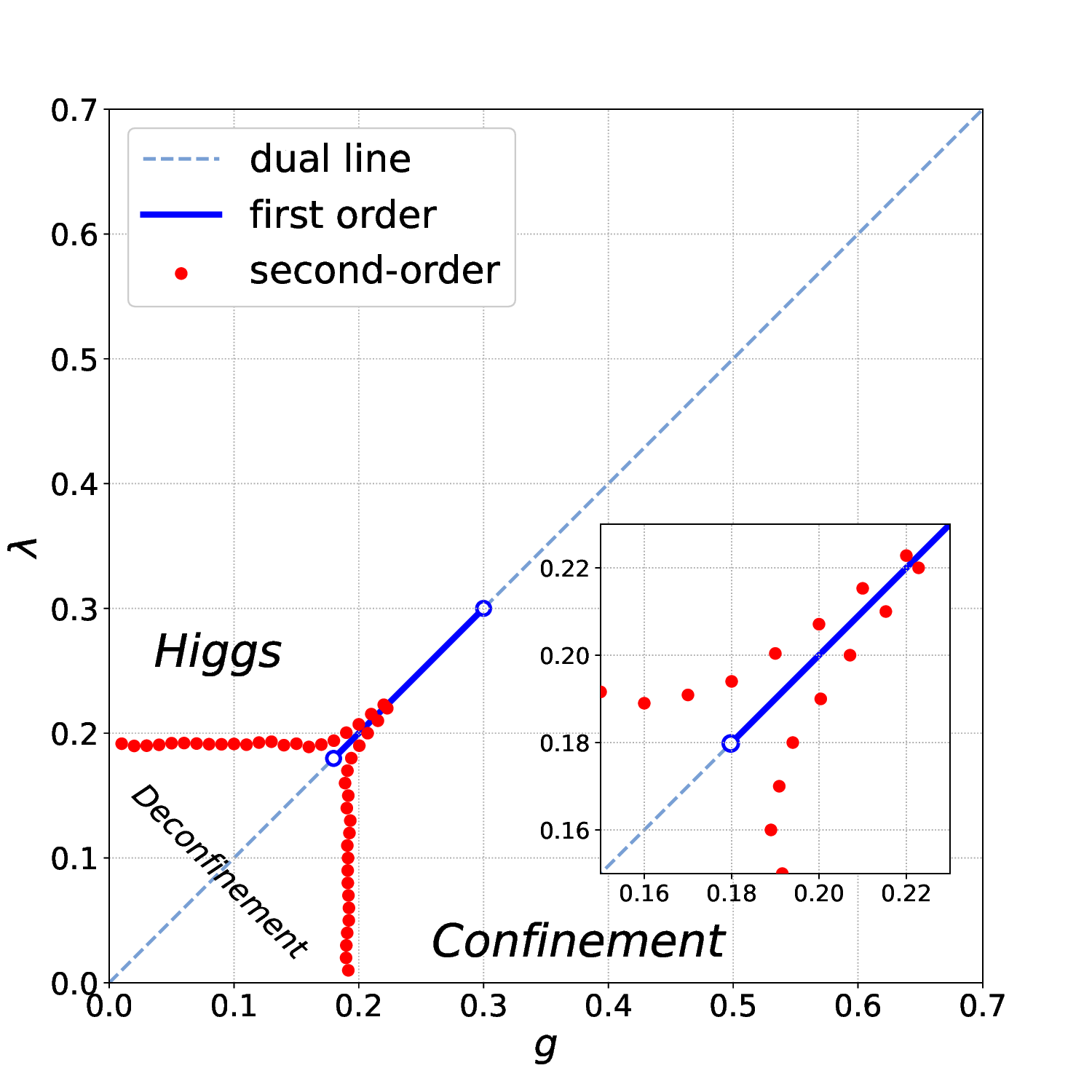}}
 \caption{\label{phasediagram}The phase diagram obtained from our calculation. The red points represent the second-order transitions and the blue line represents first-order transitions. The first-order transition line not only has a part outside the deconfined phase, but also has  a small part within  the deconfined phase.}
 \end{figure}

It has been known that in  the $\mathbb{Z}_2$  gauge-Higgs model,  there is a deconfined phase, separated from a confined phase on one hand, and from a so-called Higgs phase on the other (cf. Fig.~\ref{phasediagram}).   The phase transitions  between the deconfined  and the confined phases and between  the deconfined and the Higgs  phase are both second order, leading to a  topological region  surrounded by two second-order lines on the phase diagram. These two lines   meet at a self-dual point. The confined and Higgs phases are separated by  a finite  dual line  of  a first-order transition, beyond which the two phases are continuously connected.

However, with strong competition between matter and gauge fields, questions such as how these critical lines are connected and  where the two second-order lines meet have not been clearly answered yet, and are under debate.  It has been pointed out that there are three  possibilities~\cite{PhysRevB.82.085114}.
A quantum Monte Carlo (QMC) study provides  the evidence that the tricritical point, where the three critical lines meet,  is scale invariant and of second order \cite{2012.15845}.

In this paper, we report a  scheme  of the digital quantum simulation of the $\mathbb{Z}_2$ gauge-Higgs model.  It is digital in the sense that it  is based on Trotter decomposition of the unitary evolution \cite{trotter}. It uses  the quantum adiabatic algorithm \cite{arXiv:quant-ph/0001106} and  is implemented  in terms of quantum circuits. Given that the ground state of the toric code model~\cite{toric}  has been experimentally  prepared \cite{arXiv:2104.01180}, it is hopeful that our scheme can be realized in future experiments. Furthermore, we classically demonstrate our quantum simulation scheme using  a GPU simulator called Quantum Exact Simulation Toolkit (QuEST) \cite{QuEST} in an NVIDIA GeForce RTX 3090 GPU server. Dubbed pseudoquantum simulation, the classical demonstration of quantum simulation is also a numerical method providing useful results on this model.

We have investigated the nature of the  quantum  phase transitions, the adiabatic evolution  along the dual line on the phase diagram, as well as  the behavior near the tricritical point. Our work suggests that the tricritical point, where the two second-order lines end,  lies on the line of the first-order transition, but not at  the lower end of it (see  Fig.~\ref{phasediagram}).

The rest of the paper is arranged as follows: In Sec. \ref{sec:ModDes},  the  $\mathbb{Z}_2$ gauge -Higgs model is briefly introduced. In Sec. \ref{sec:Algo}, we  elaborate on the  digital quantum adiabatic algorithm and its realization in terms of   quantum circuits. In Sec. \ref{sec:topo}, we describe the preparation of the initial state, as well as  the topological phase. In Sec. \ref{sec:Result}, we discuss the critical points according to the analysis of the density of states (DOS). The  Trotter error is analyzed in Sec. \ref{sec:trotter}, with some details given in the Appendix. We make a comparison between our method and exact diagonalization   in Sec. \ref{sec:cmp}. A  summary is made in Sec. \ref{sec:sum}.

\section{Model description}
\label{sec:ModDes}

The Hamiltonian of  the $\mathbb{Z}_2$ gauge-H iggs model is
\begin{equation}\label{eq:GH}
\begin{split}
	H = -J\sum_v\tau^x_v - h\sum_p B_p^z
    - g\sum_l\sigma^x_l - \lambda \sum_{\langle j,k\rangle}
    \tau^z_j \sigma^z_{\langle j,k\rangle}\tau_k^z,
    \end{split}
\end{equation}
where $\sigma$ denotes the gauge field defined on the links, $\tau$ denotes the Ising matter field defined at the vertices, and
\begin{equation}
	B^z_p=\prod_{j \in \partial p}\sigma^z_j,
\end{equation}
is the tensor product of four $\sigma^z$'s on the sides of a plaquette $p$. Gauss's law requires that the ground state be invariant under the action of
\begin{equation}\label{gauss}
Q_{v}^x \equiv \tau^{x}_{v} A_v^x,
\end{equation}
with  $ A_v^x \equiv  \prod_{j\in \partial v} \sigma^{x}_{j}$.

Under a mathematical mapping, the $\mathbb{Z}_2$ guage-Higgs model  as given in Eq. (\ref{eq:GH}) is equivalent to the  toric code model in two transverse fields \cite{PhysRevB.82.085114},  which is thus studied  in this paper. We focus on the parameter subspace with  $J=h=1/2$. Hence the Hamiltonian reads
\begin{equation}\label{eq:effi}
	H = -\frac{1}{2}\sum_vA_v^x - \frac{1}{2}\sum_p B_p^z
    - g\sum_l\sigma^x_l - \lambda\sum_l\sigma^z_l.
\end{equation}

The total number of qubits is 19, with one for the ancilla and 18 for a $3\times 3$ lattice model on the torus. The size is small. Unfortunately it is very difficult to make it larger.  A $4\times 4$ lattice would need 33 qubits, which is too large  an  increase for both the power of present classical computation and  real quantum simulation in present quantum hardware, let alone even larger lattice size for the purpose of  finite size scaling.

The main goal of our work is to present and classically demonstrate the scheme of the digital quantum simulation, while the  calculations are a proof-of-principle demonstration  shedding some light on the nature of   the quantum phase transitions in this model.

\section{Quantum Algorithm }
\label{sec:Algo}

The purpose is to obtain the energy of the system as a function of  the parameters  $\lambda$ and $g$.  Since there exists the self-duality, we only need to investigate the behavior below the self-dual line $\lambda=g$ on  the $\lambda$-$g$ parameter plane.   The $g$ axis, where $\lambda=0$,  represents the pure $\mathbb{Z}_2$ gauge theory,  for which  we  prepare the initial ground state on a point   $P(p,0)$ on the $g$ axis.

\begin{figure}[htb] 
 \center{\includegraphics[width=8cm]  {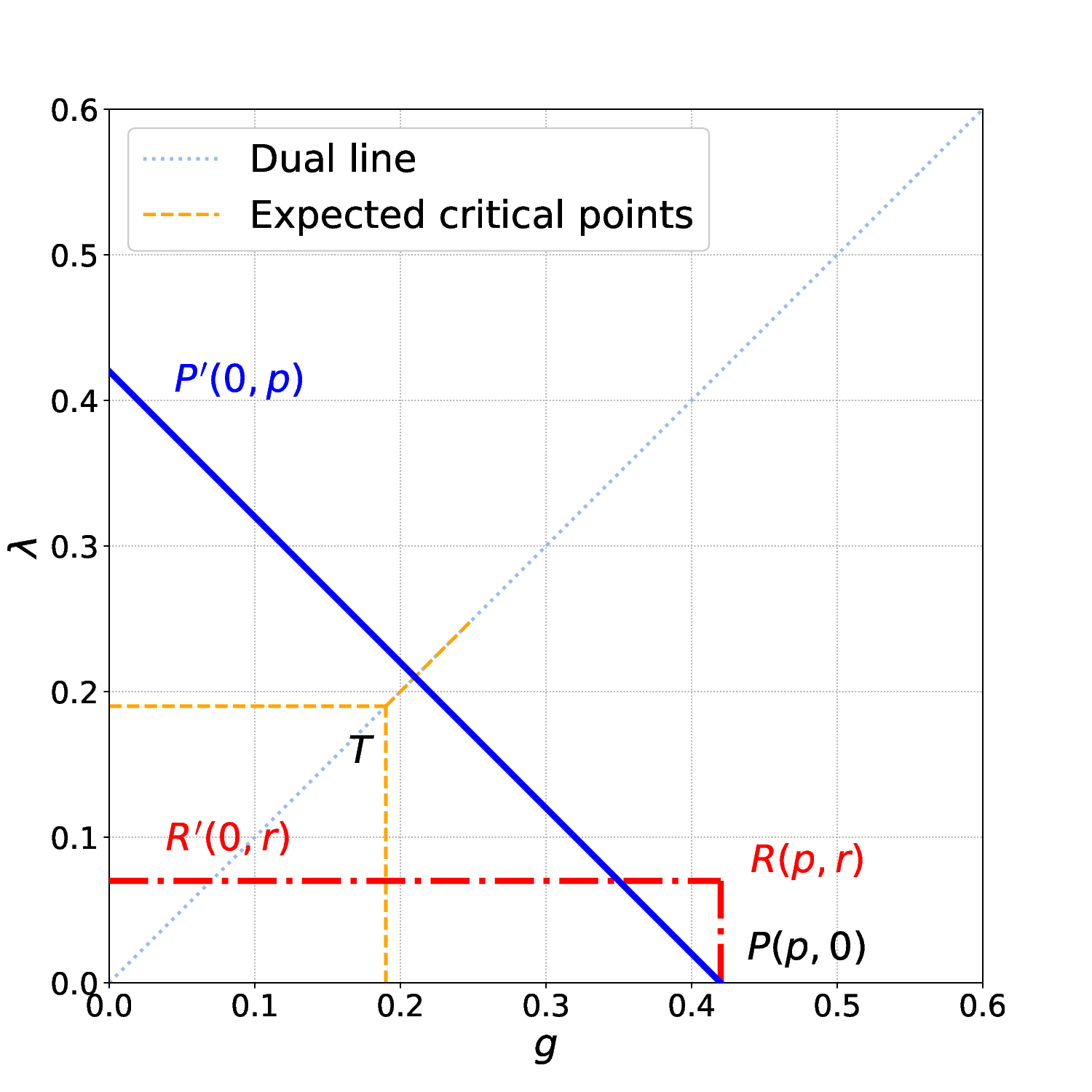}}
 \caption{\label{evo}Two paths of parameter variation in our simulations.}
 \end{figure}

Two paths of parameter variation  are used for the adiabatic algorithm. As shown  in Fig. \ref{evo}, on a path depicted as a broken red line, the parameters  vary first from $P(p,0)$ to $R(p,r)$, then to $R'(0,r)$. On a path depicted as a solid blue line, the parameters vary from $P(p,0)$ to $P'(0,p)$ on the straight line.

Now, we introduce the algorithm for the digital quantum simulation, which is  implemented in terms of quantum circuits  and uses a quantum adiabatic algorithm to evolve the ground state along paths in the parameter space.  For convenience,  we write the Hamiltonian  as
\begin{equation}
H=H_1+H_2+H_3,
\end{equation}
where
\begin{equation}
\begin{split}
&H_1 = -g\sum_l\sigma^x_l,\\
        &H_2 = -\frac{1}{2}\sum_vA_v^x
        - \frac{1}{2}\sum_p B_p^z, \\
        &H_3 = -\lambda\sum_l\sigma^z_l
        \end{split}.
\end{equation}

We decompose the evolution operator $e^{-iHt}$ by using the second-order Trotter-Suzuki formula \cite{Hatano2005},
\begin{equation}
\begin{split}
e^{-iHt} \approx e^{-iH_1\frac{t}{2}} e^{-iH_2\frac{t}{2}} e^{-iH_3t} e^{-iH_2\frac{t}{2}} e^{-iH_1\frac{t}{2}},
\end{split}
\end{equation}
where
\begin{equation}\label{decom}
\begin{split}
&e^{-iH_1\frac{t}{2}}=
\prod_le^{ig\sigma^x_l\frac{t}{2}},\\
        &e^{-iH_2\frac{t}{2}}=\prod_ve^{i\frac{1}{2}A^x_v\frac{t}{2}}
        \prod_pe^{i\frac{1}{2}B^z_p\frac{t}{2}},\\
        &e^{-iH_3t}=\prod_le^{i\lambda\sigma^z_lt}.\\
       \end{split}
\end{equation}
The decompositions in Eq. (\ref{decom})  do not  generate  errors,  since the summands in $H_j$ commute with each other.

$e^{-iH_1\frac{t}{2}}$ and $e^{-iH_3{t}}$ can be realized simply by using  the rotation gates $\mathrm{R}_x$ and $\mathrm{R}_z$, respectively:
\begin{equation}
\label{gate1}
\begin{split}
&\prod_l e^{ig\sigma^x_l\frac{t}{2}} = \prod_l \mathrm{R}_x^l(-gt),\\		
&\prod_l e^{i\lambda\sigma^z_l{t}} = \prod_l \mathrm{R}_z^l(-2\lambda t).\\
\end{split}
\end{equation}

By introducing an ancilla $a$,   $e^{i\frac{1}{2}B_p^z\frac{t}{2}}$ can be realized as
\begin{equation}
\begin{split}
&e^{i\frac{1}{2}B_p^z\frac{t}{2}} = \mathrm{U}_1 [\mathrm{R}_z(-\frac{t}{2})]_a \mathrm{U}_1^{\dagger},
\end{split}
\end{equation}
where
\begin{equation}
\begin{split}
\mathrm{U}_1 = \prod_{j\in p}\mathrm{CNOT}_{j\to a}.
\end{split}
\end{equation}

$e^{i\frac{1}{2}A_v^x\frac{t}{2}}$ can be realized similarly, and we only need  four additional  Hadamard gates to switch into $z$ basis the four spins on the sides connected at $v$—that is,
\begin{equation}
\begin{split}
&e^{i\frac{1}{2}A_v^x\frac{t}{2}} = \mathrm{U_2}^{\dagger}[\mathrm{R}_z(-\frac{t}{2})] \mathrm{U_2},\\
\end{split}
\end{equation}
where
\begin{equation}
\begin{split}
\mathrm{U_2} = [\prod_{k\in v}\mathrm{CNOT}_{k\to a} ][\prod_{k\in v}{\mathrm{H}}_k].\\
\end{split}
\end{equation}
Note that we use the convention for the time order of   the operators—that is,  from right to left. We omit drawing the circuits, which is straightforward.

To ensure adiabaticity in the variation of the parameters, evolution on each path is divided into numerous tiny steps, each with the same duration of time.

\section{Preparation of initial state }
\label{sec:topo}

The initial ground state at the parameter point  $P(p,0)$ is the same as an intermediate state in our previous work on the pure $\mathbb{Z}_2$ gauge theory  \cite{JHEP08(2020)160}, as the mere addition of $A_v$, which commutes with $B_p^z$, does not  change  the state. As we have mentioned before, this has been  experimentally realized \cite{arXiv:2104.01180}. In our demonstration in the classical simulator, nevertheless, for convenience, we can simply use projections to prepare such a state, although it is not convenient in experiments.

The ground state is prepared  from $|00\cdots00\rangle$, by using
\begin{equation}\label{ini1}
\prod_p \{   [\prod_{j\in p} \mathrm{CNOT}_{j\to a}] [\mathrm{P}_{a=|0\rangle}]
  [\prod_{j\in p} \mathrm{CNOT}_{j\to a}]           \}   \prod_l H_l,
\end{equation}
where P represents the projection of  the ancilla to be $|0\rangle$. The idea is  to first  generate  the equal  superposition of all the computational basis  states,  by  using a  Hadamard gate   on each  qubit, then for each plaquette, to apply four {\scshape cnot} gates to transfer the  information of each basis state to an ancilla \cite{JHEP08(2020)160}.  It leads to  a superposition of all possible configurations with  $B_p=1$ for every $p$. This  is a ground state at parameter point $O(0,0)$ on $\lambda$-$g$ plane, denoted as  $\psi_1$.

Then we adiabatically evolve $\psi_1$ along the $g$ axis toward $P(p,0)$, by using the circuit-based digital quantum adiabatic algorithm in Sec.~\ref{sec:Algo}.

If we initially prepare a ground state at $P'(0,p')$ by adiabatic evolution from the one at $O(0,0)$,    the state prepared at $O(0,0)$ should be the superposition of all configurations with $A_v=1$ for every $v$.

The quantum circuit for preparing $\psi_1'$ is
\begin{equation}
\label{ini2}
\prod_v \{ [\prod_{k\in v} \mathrm{H}_k \prod_{k\in v}
    \mathrm{CNOT}_{k\to a}    ][\mathrm{P}_{a=|0\rangle}][
    \prod_{k\in v} \mathrm{CNOT}_{k\to a}\prod_{k\in v} \mathrm{H}_k ] \} .
\end{equation}

The ground states  can be described with the aid of  't Hooft loop operators
$V_{\mu}^x$ and noncontractible Wilson loop operators  $W_{\nu}^z$, and the eigenstates of $V_{\mu}^x$ on the $g$ axis are dual to the eigenstates of $W_{\nu}^z$ on the $\lambda$ axis \cite{FradkinBook}. $\psi_1$,  $\psi_2=W_1\psi_1$, $\psi_3=W_2\psi_1$ and $\psi_4=W_1W_2\psi_1$ are different eigenstates of $V_{1}^x$ and $V_{2}^x$, with eigenvalues $(\pm 1,\pm 1)$. As $V_{1}^x$ and $V_{2}^x$ are  topological, $\psi_1$,  $\psi_2$, $\psi_3$, and $\psi_4$ are in different topological sectors, and a state in one topological sector cannot evolve into other topological sectors. Moreover,  $\psi_1$,  $\psi_2$, $\psi_3$ and $\psi_4$ are four  degenerate ground states at $O(0,0)$.

It turns out that
$\psi_1' = (\psi_1 + \psi_2 + \psi_3 +\psi_4)/2$.
Therefore, if the evolution is restricted in a topological phase,  $\psi_1$ and $\psi_1'$   cannot evolve to each other, because of topological protection. Hence  the evolution from $P(p,0)$ to $P'(0,p')$ should be different from the evolution from  $P'(0,p')$ to $P(p,0)$, on the path depicted as a solid blue line in Fig. \ref{evo}.

\begin{figure}[htbp]
\centering
 {
\includegraphics[width=6cm]{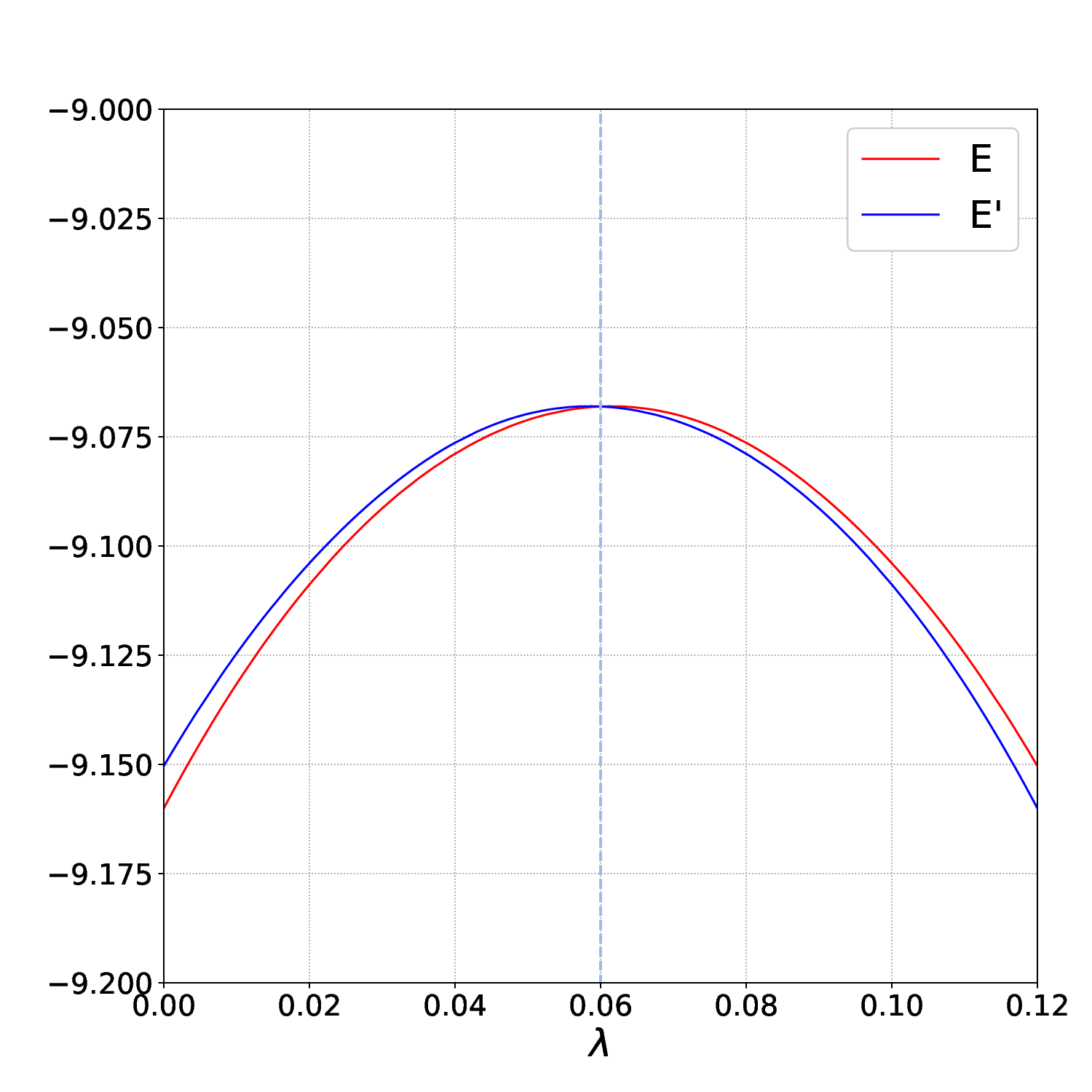}
}
 {
\includegraphics[width=6cm]{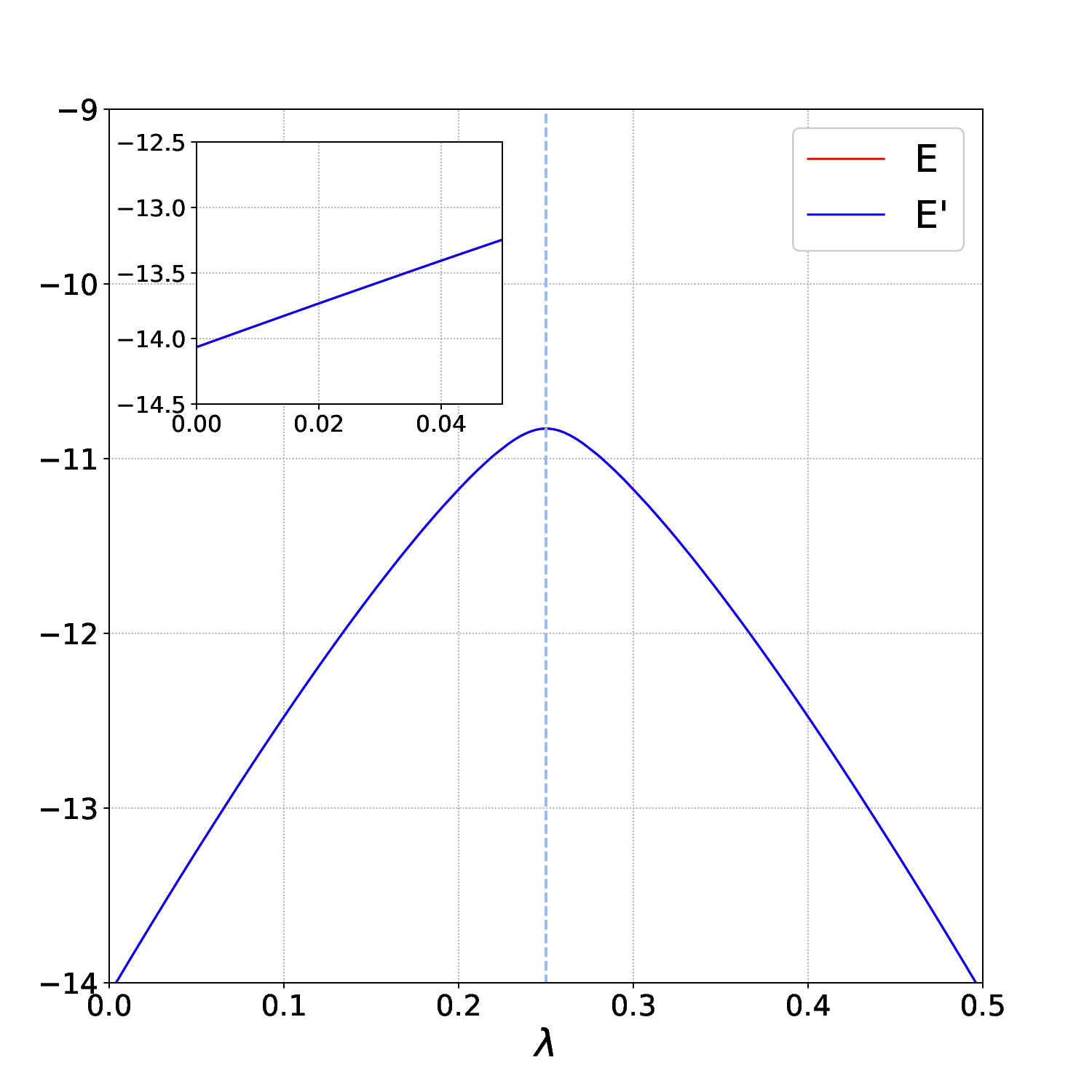}
}
\caption{\label{topo}Energy  as a function of $\lambda$ along the blue path. (a) For $p=p'=0.12$, $E=-9.16003$ while $E'=-9.15039$, where $E$ is the energy of the state evolved from that prepared at $P$, while $E'$ is the energy of the state prepared at $P'$. Two energy curves cross, indicating  a topological phase. (b) For $p=p'=0.50$, $E'=-14.06550$ and $E=-14.06553$. Two energy curves nearly coincide and the energy  is symmetric with respect to the center of the blue path, indicating a nontopological phase.}
\end{figure}

We prepare each of these two ground states on the corresponding parameter point and evolve it  to  the other parameter point  along the solid
blue path in Fig. \ref{evo}. The result shows that in Fig. \ref{topo}, representing the functions of the parameter $\lambda$, the two  curves cross   if the path is chosen within the deconfined phase, and they do  not cross if the path is  outside  of  the deconfined phase.
The crossing here  is an indication of topological phase, and it shows the irreversibility of  the adiabatic evolution, which is history dependent. This feature cannot show up in exact diagonalization or QMC, which directly give results at each parameter point.
Note that the crossing here is not the level crossing in a finite lattice that becomes an avoided level crossing in an infinite lattice.

\section{Phase diagram}
\label{sec:Result}

\begin{figure*}[htbp]
\centering
 {
\includegraphics[width=5cm]{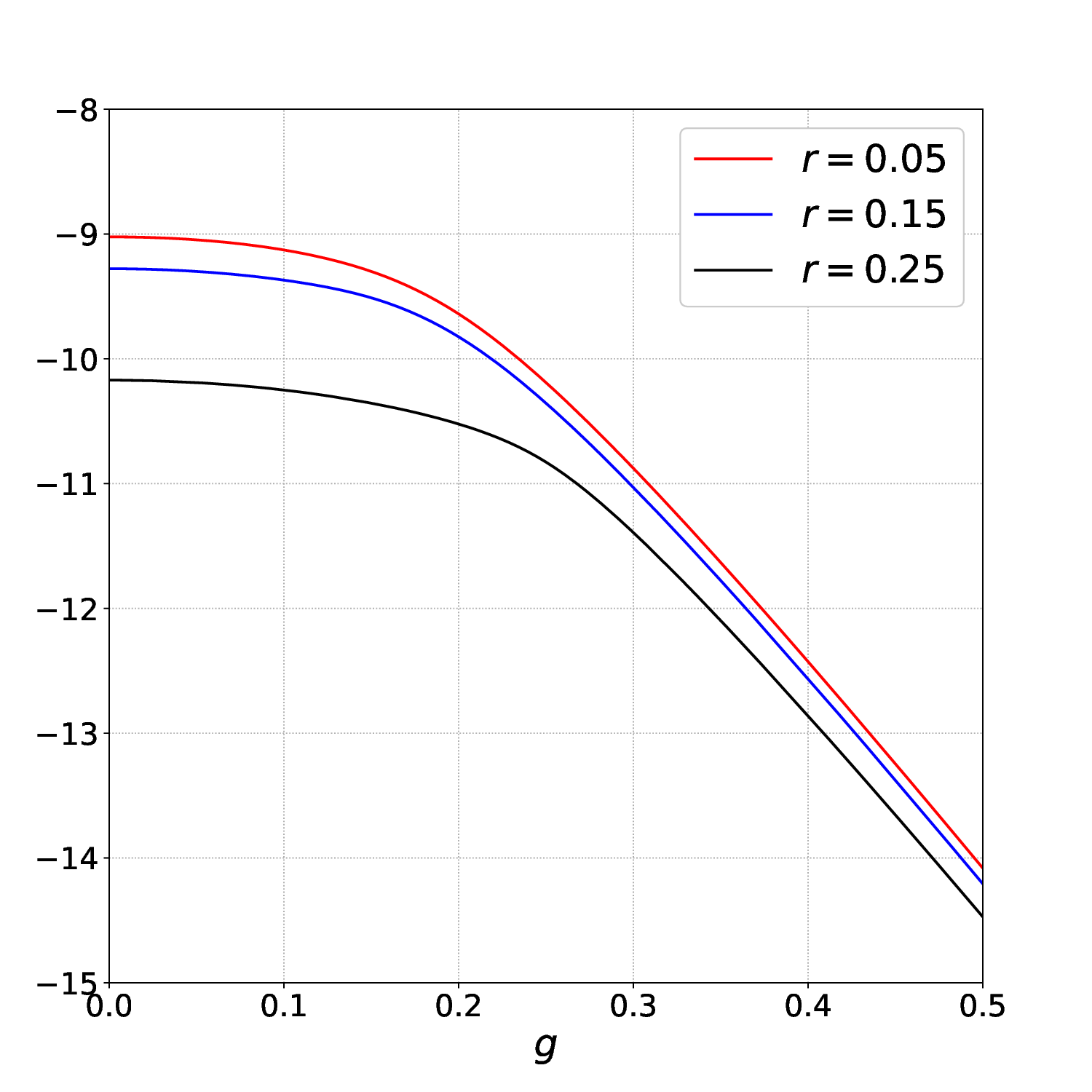}
}
 {
\includegraphics[width=5cm]{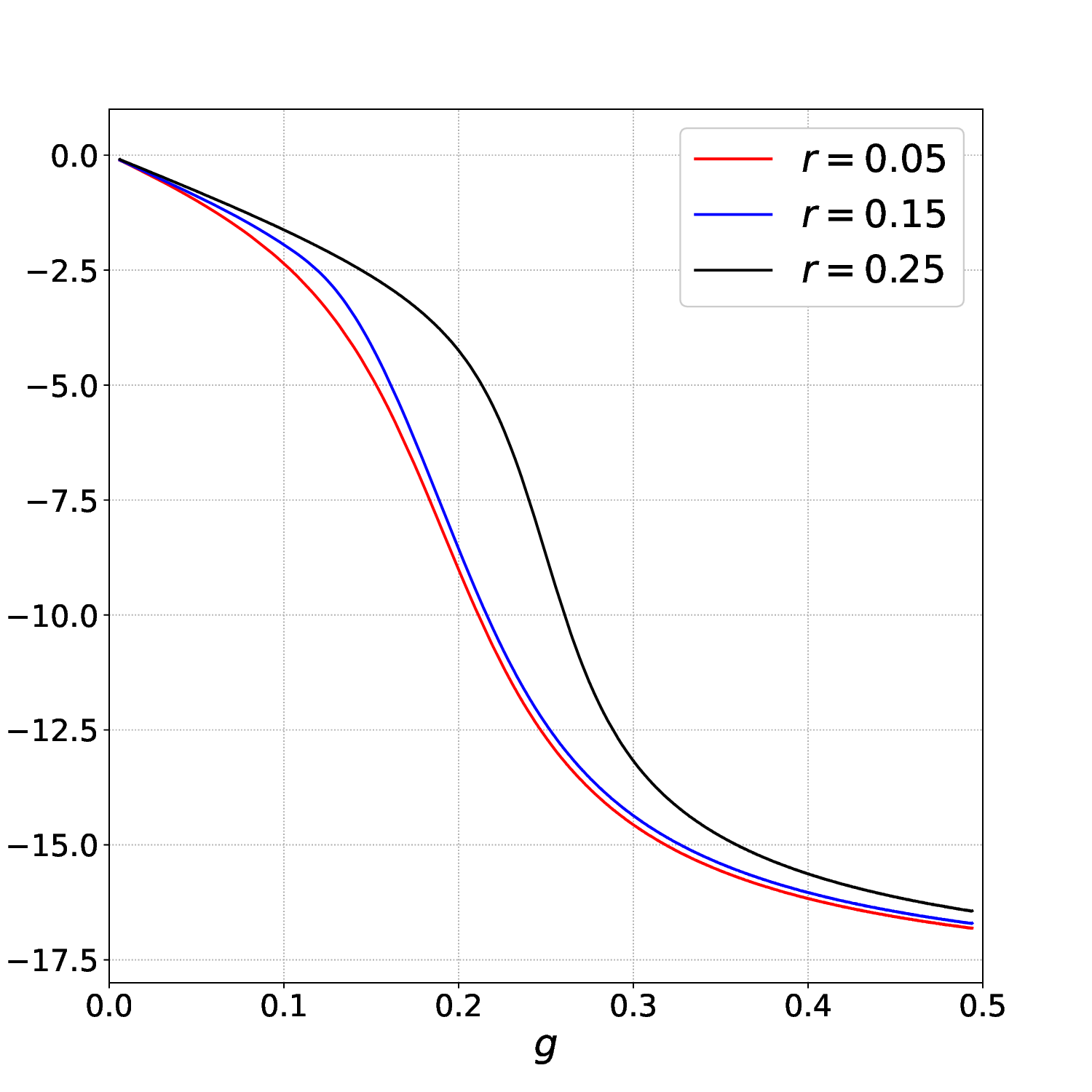}
}
 {
\includegraphics[width=5cm]{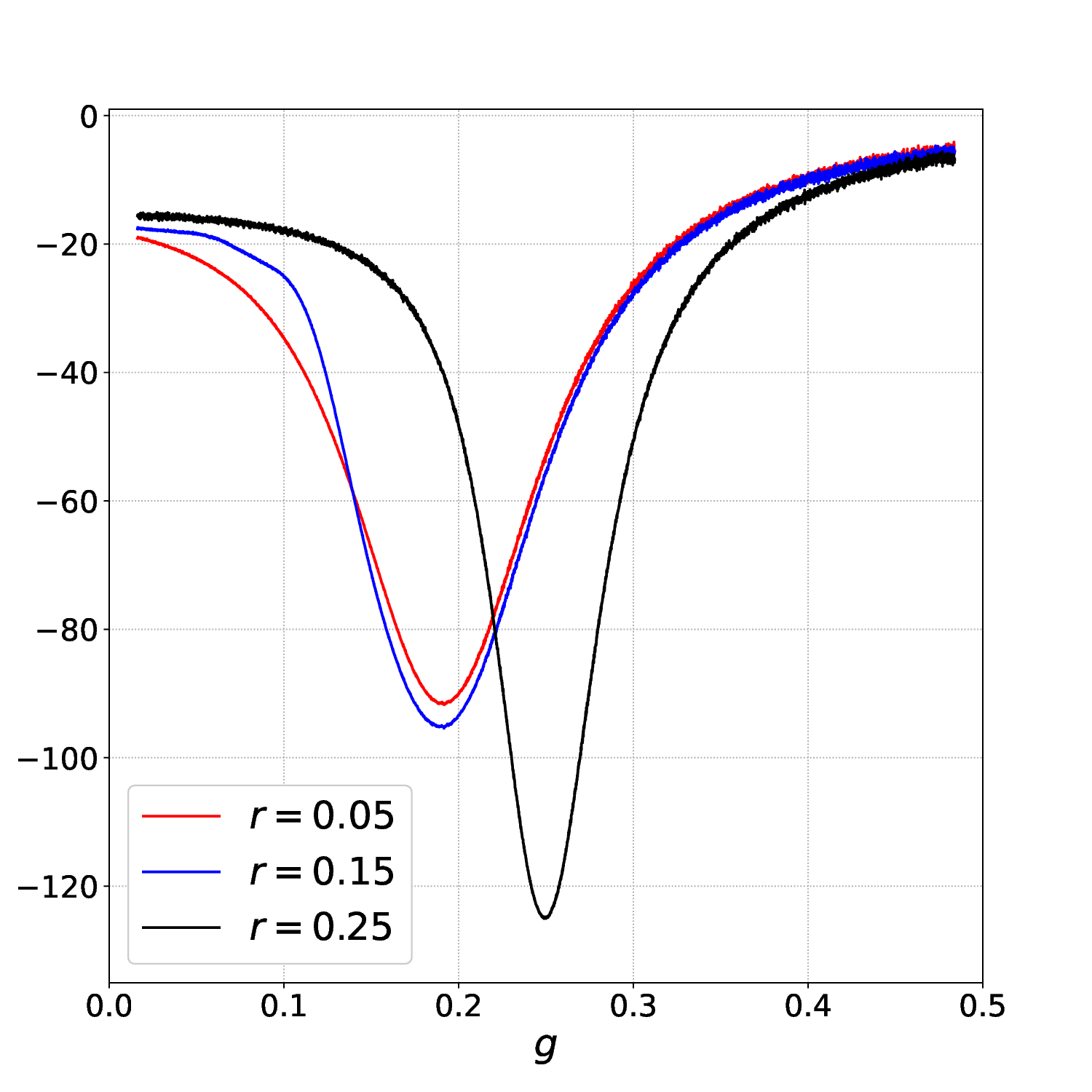}
}
\caption{\label{sec_order} On the evolution paths in  broken red lines, with $\lambda$ being   $r=0.05,0.15,0.25$ (see Fig.~\ref{evo}), the plots show (a) energy,
(b) the first derivative of the energy, and (c) the second derivative of the energy, as functions of $g$.
 The minima in (c) are at $g=0.1920$, $0.1916$, and $0.2489$ respectively.}
\end{figure*}

We now look for the fingerprints  of the critical points of quantum phase transitions as the extremal points in the second derivatives of the energy.

After numerous trials, we choose $p=0.5$ for the path depicted as the broken red line in Fig.~\ref{evo}, of which the $\lambda$ value is denoted as  $r$. The result suggests that for $r<0.19$, the critical value of $g$ is $g_c\approx 0.19$, while for $r>0.19$, $g_c \approx r$, as depicted in  Fig. \ref{sec_order}.
In this way, we find that the two critical lines surrounding the deconfined phase meet at $g\approx0.24$, as shown in  Fig. \ref{phasediagram}.

\begin{figure}[htbp]
\centering
 {
\includegraphics[width=6cm]{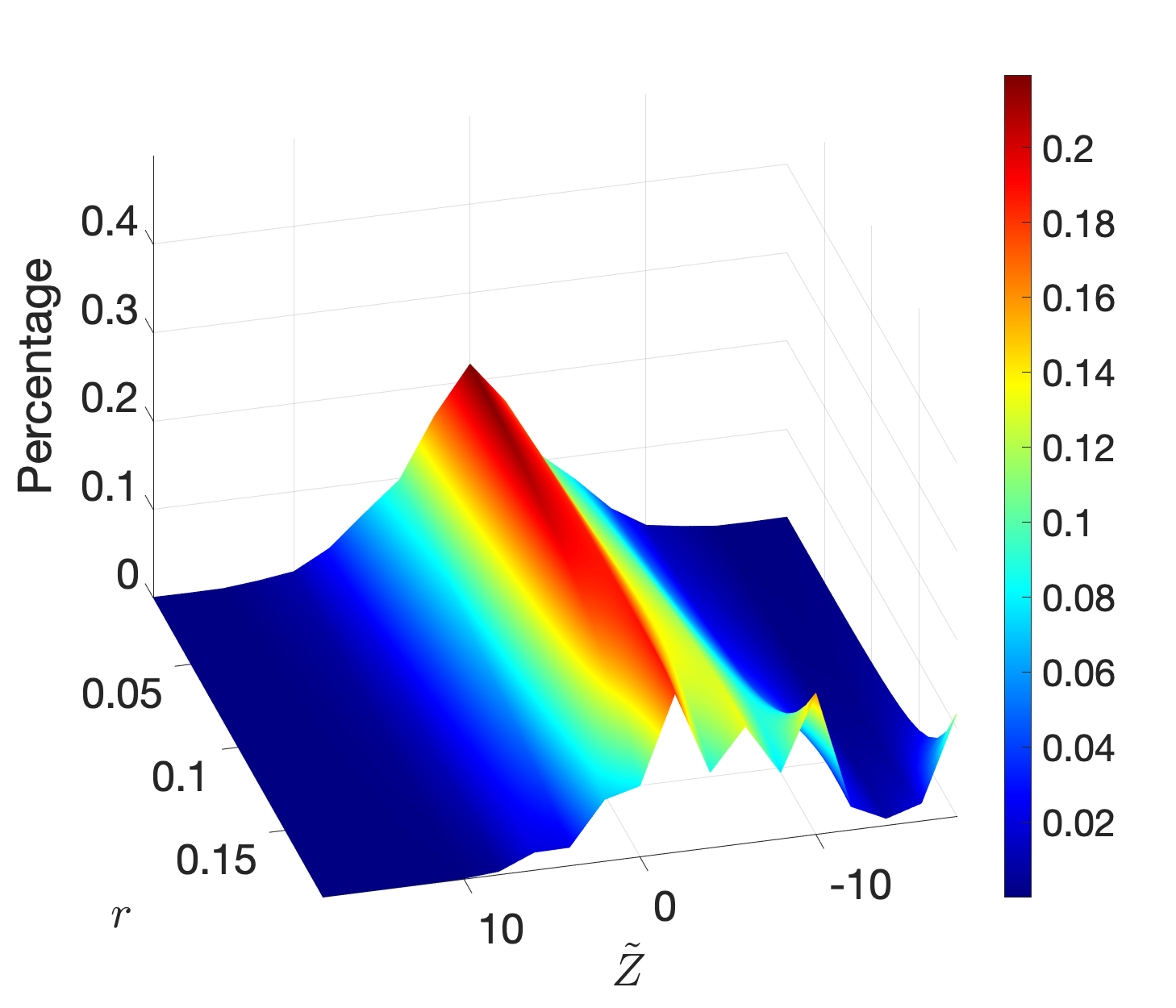}
}
 {
\includegraphics[width=6cm]{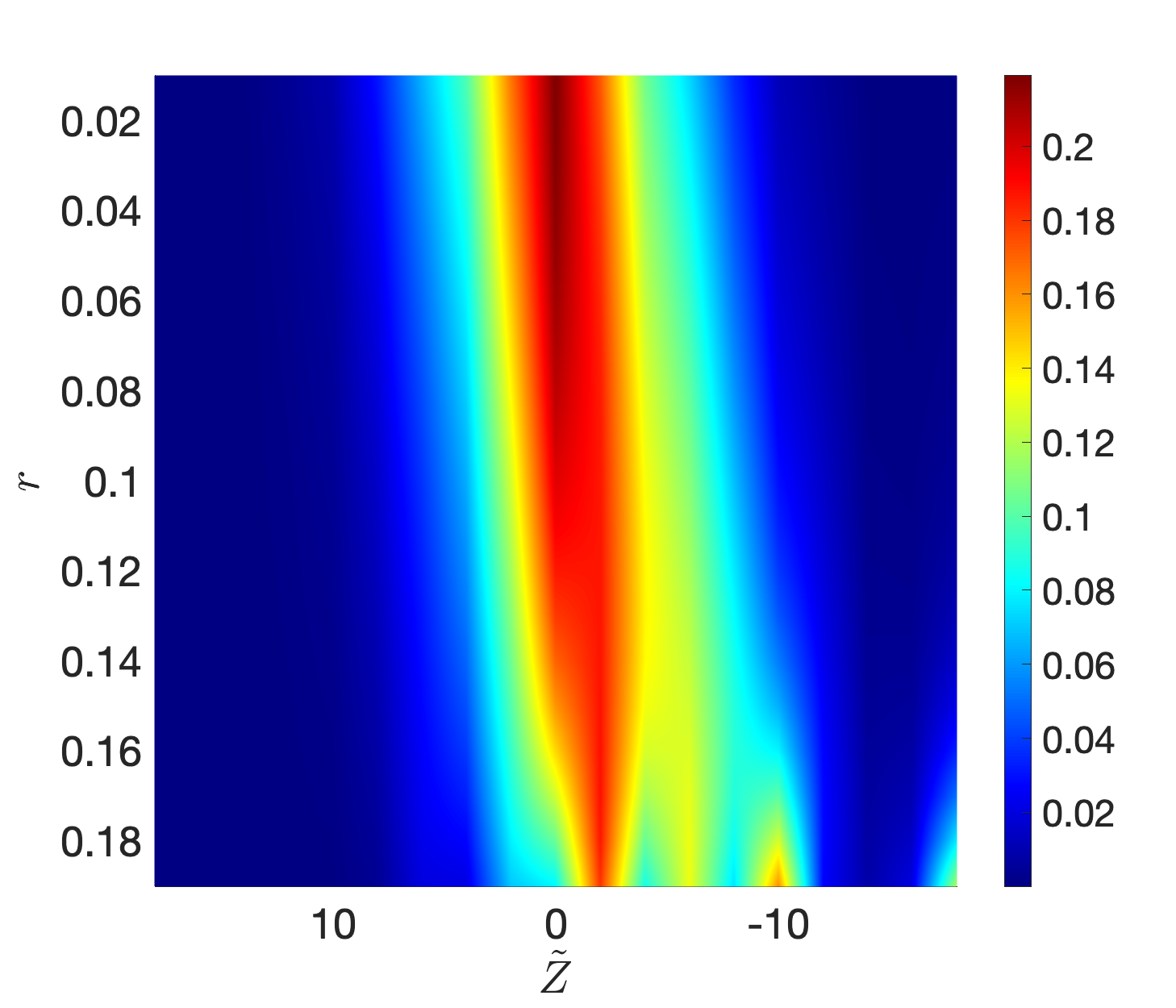}
}

 {
\includegraphics[width=6cm]{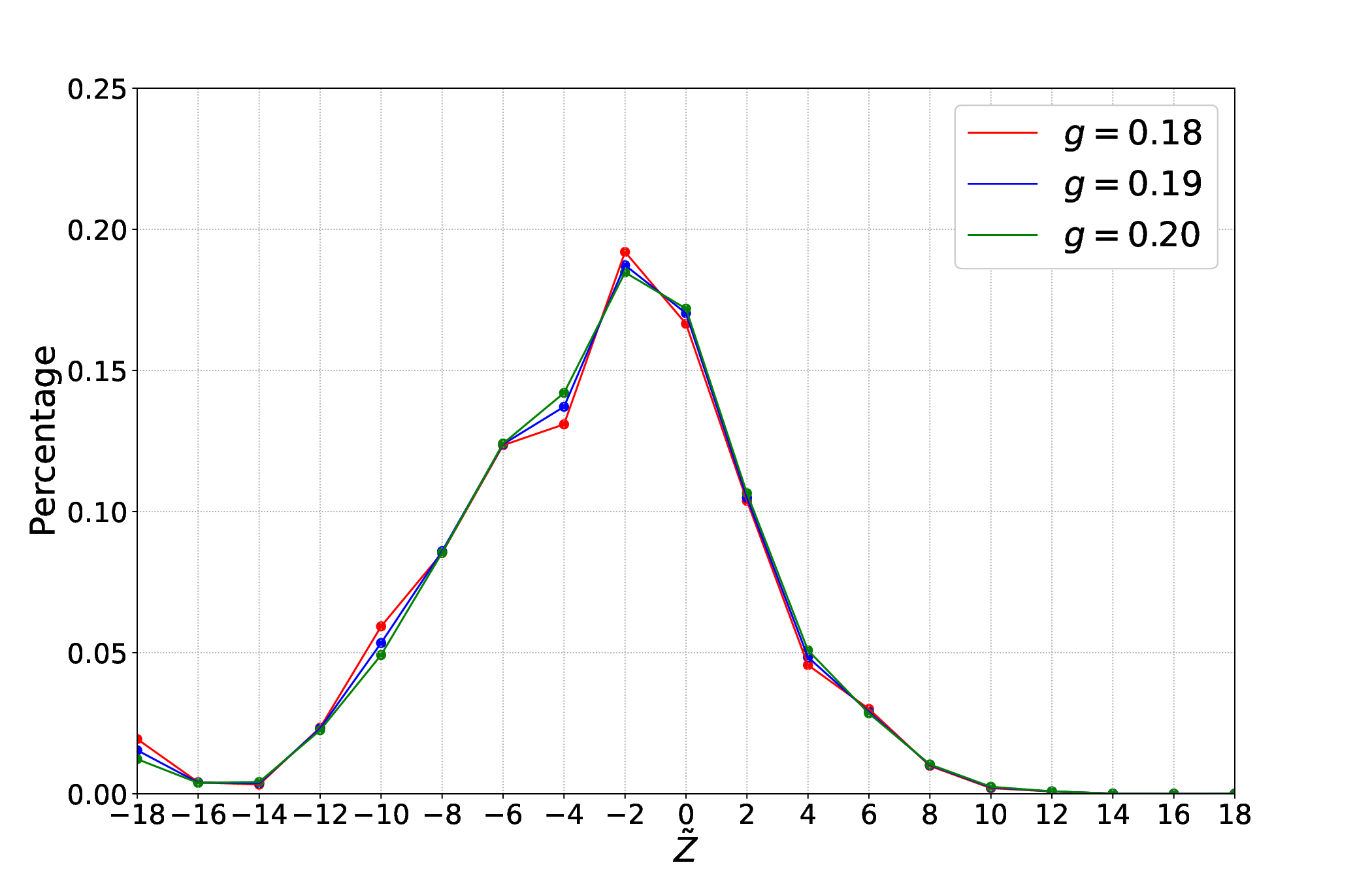}
}
 {
\includegraphics[width=6cm]{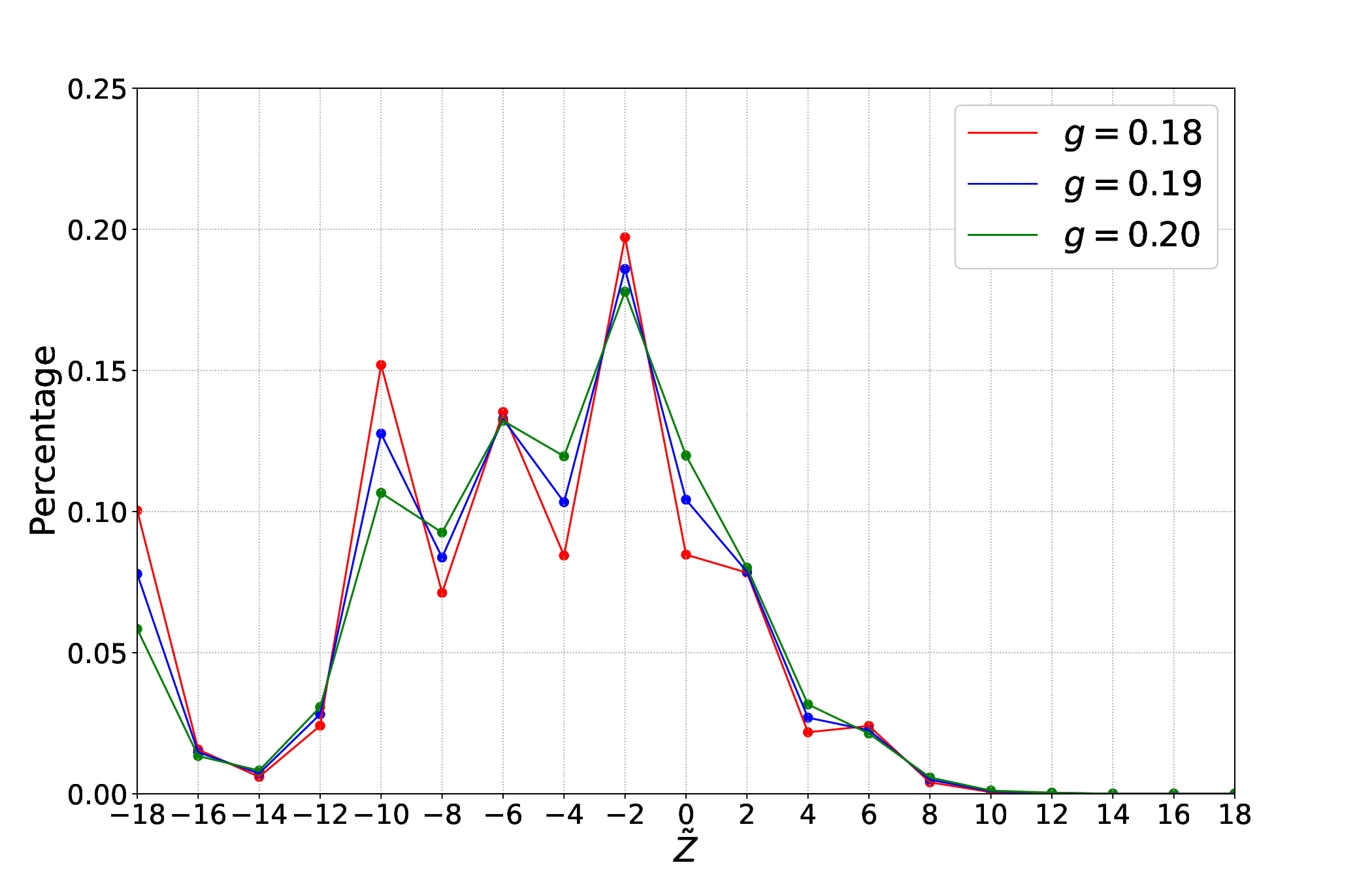}
}
\caption{ \label{dos_sec} (a) (b) DOS of $\tilde{Z}$ for $g=0.19$ and $r\in[0.01,0.19]$ on the red paths.
(c) DOS of $\tilde{Z}$ for $r=0.14$. (d) DOS of $\tilde{Z}$ for $r=0.18$.}

\end{figure}

As shown in Fig.~\ref{dos_sec},  we analyze the order of the quantum  phase transition by considering  DOS (density of states), where the states refer to the eigenstates of  $\tilde{Z}=-\sum_l\sigma^z_l$, as  in our previous work.  When $r<0.16$, the DOS of $\tilde{Z}$ exhibits only one maximum in each phase.   This is similar to the case of the pure $\mathbb{Z}_2$ gauge theory. Hence the quantum phase transition here is of  second order when $r<0.16$. This verifies that the two  critical lines surrounding the deconfined phase are second order.

DOS becomes more and more disordered when $r>0.16$,  which represents the region close to the meeting point  $T$ of the two second-order critical lines. Similar phenomena can also be observed on the paths in solid blue lines.

\begin{figure}
\centering
 {
\includegraphics[width=6cm]{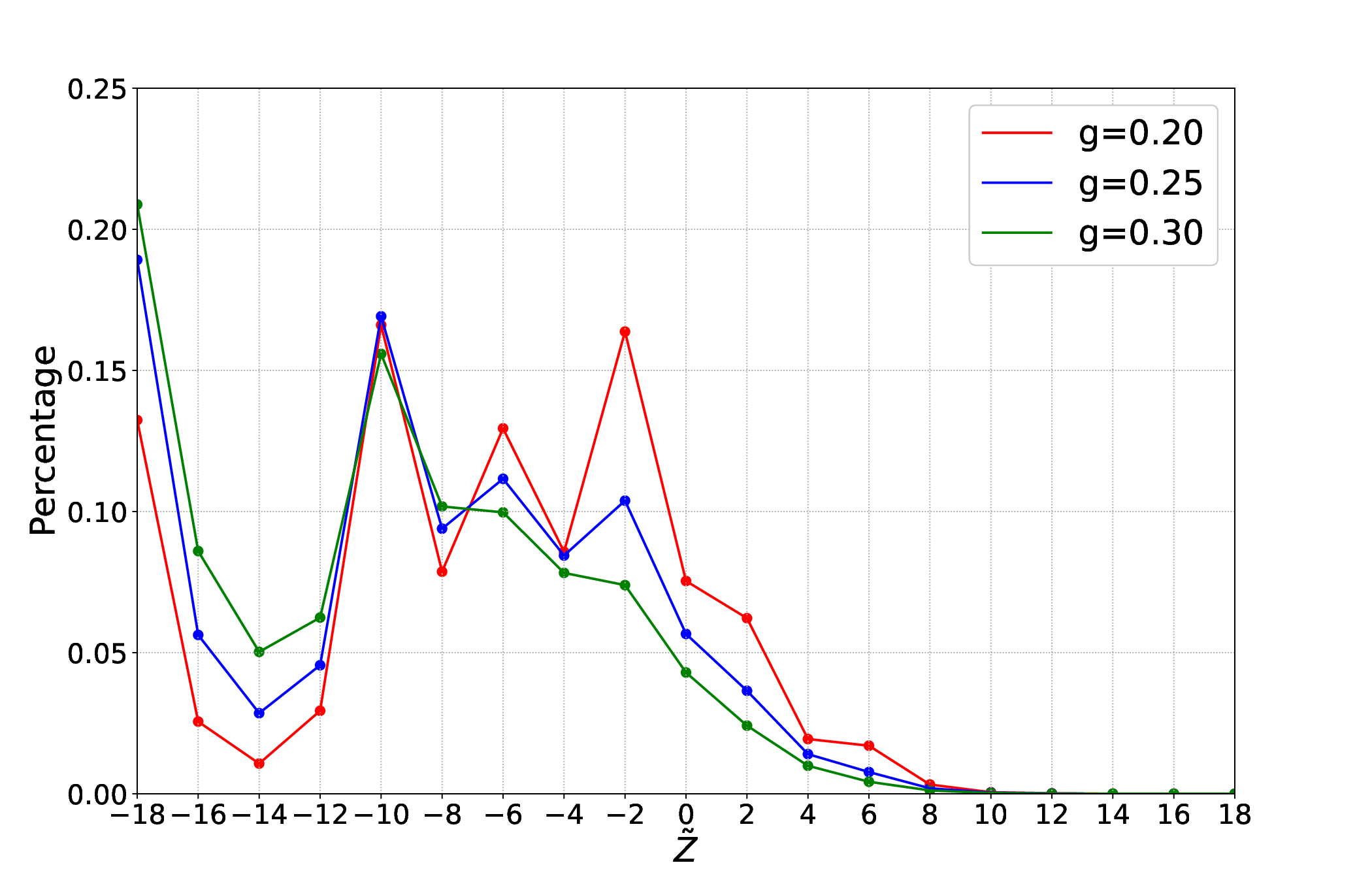}
}
 {
\includegraphics[width=6cm]{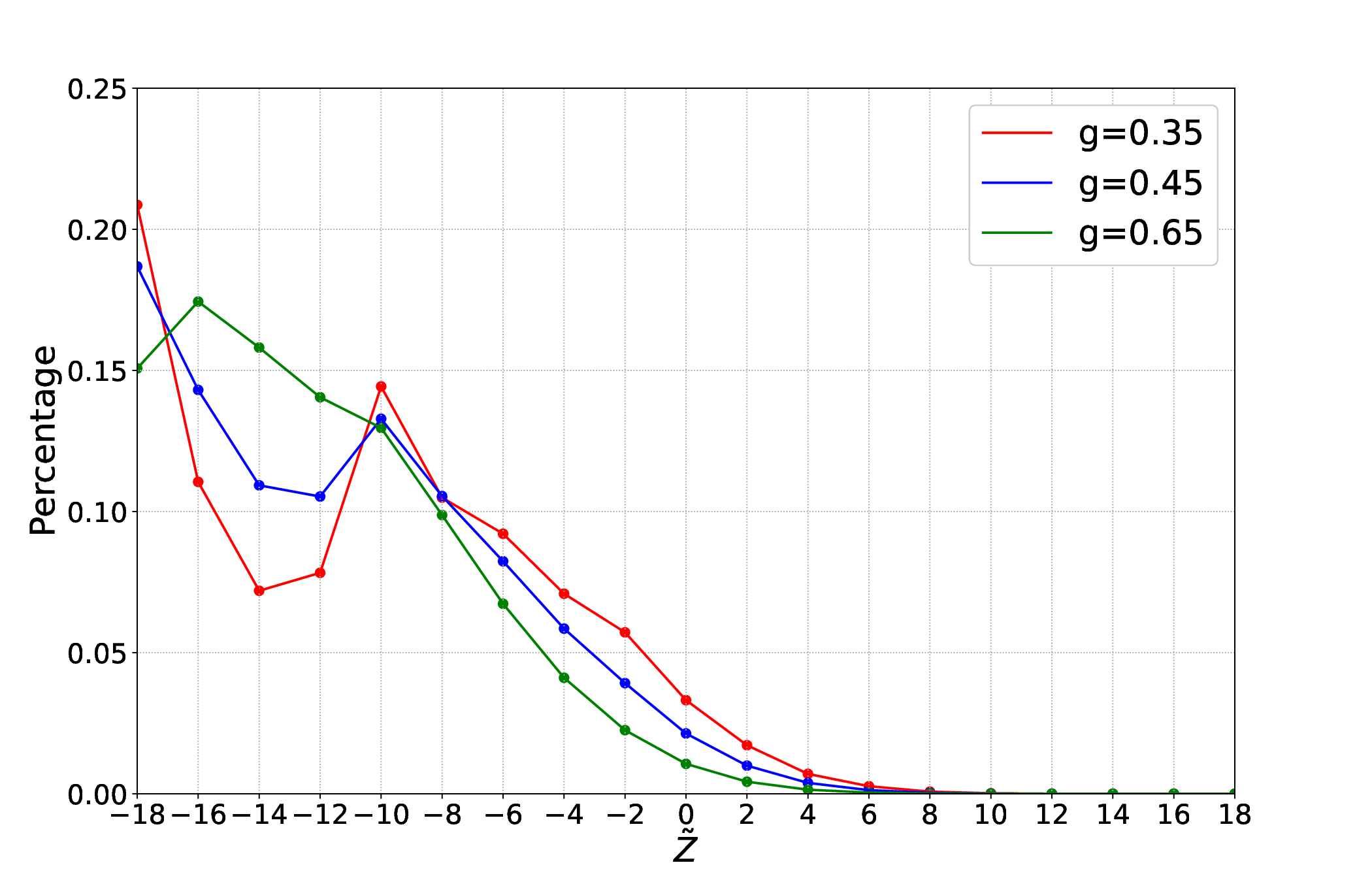}
}
\caption{\label{dos_dual} DOS of $\tilde{Z}$ on the dual line. The disorder recedes as $g$ increases. (a) At $g\approx 0.30$, the multimaxima are reduced to  two maxima. (b) At $g\approx 0.65$, we detect a single maximum.}
\end{figure}

A characteristic feature of the first-order  phase transition is the coexistence of different phases. As shown in Fig. \ref{dos_dual}, The conversion from single-maximal features to multimaximal features of the DOS of $\tilde{Z}$  can be  regarded as suggesting  the conversion from a second-order phase transition to a first-order phase transition.  However, as our system is small, although
the multimaximal feature appears in DOS when $r>0.16$ in red paths, the first-order phase transition does not necessarily  appear when  $0.16<g<0.24$. It is likely that it appears only after the two second-order lines cross at $g \approx 0.24$.

\begin{figure}[htbp]
\centering
 {
\includegraphics[width=6cm]{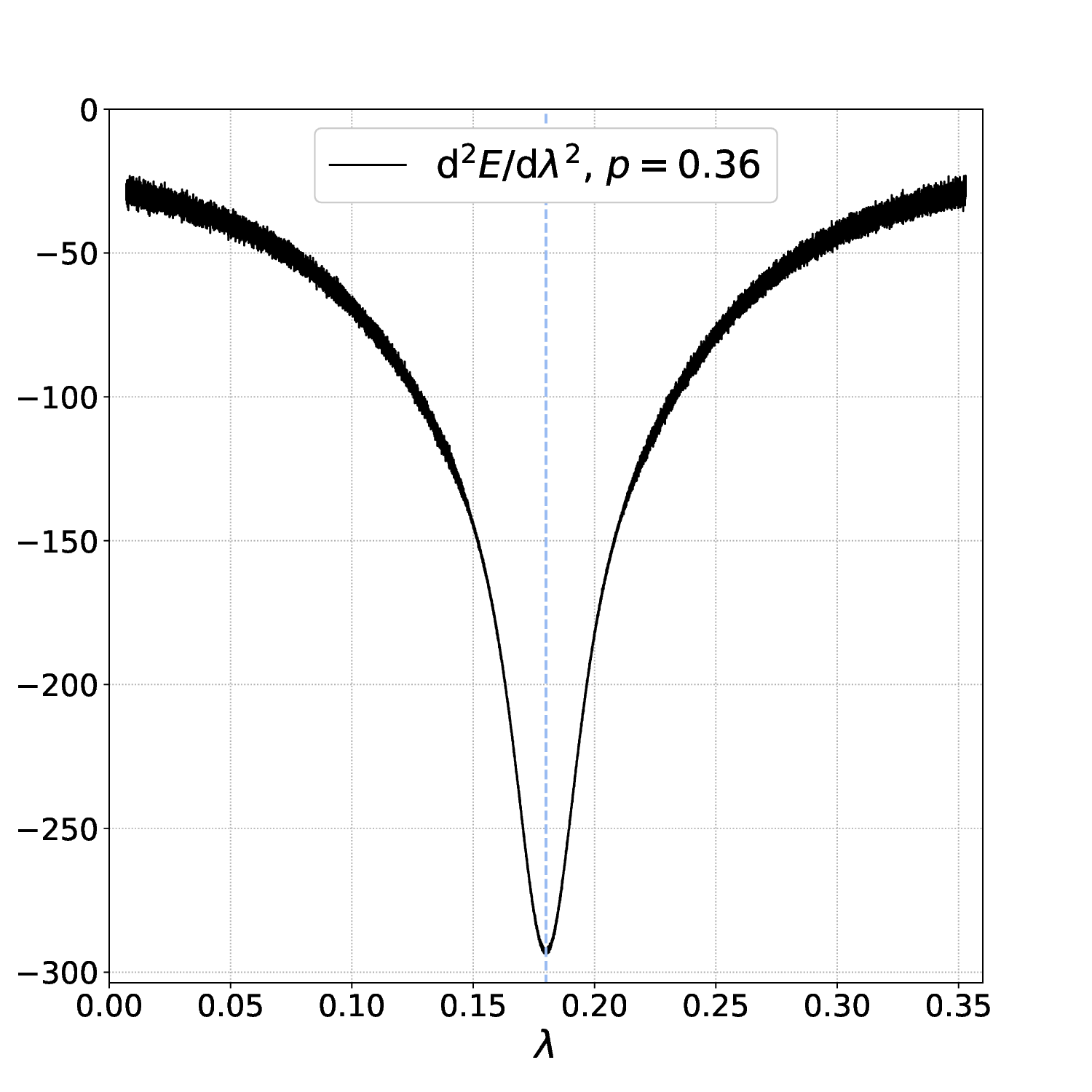}
}
 {
\includegraphics[width=6cm]{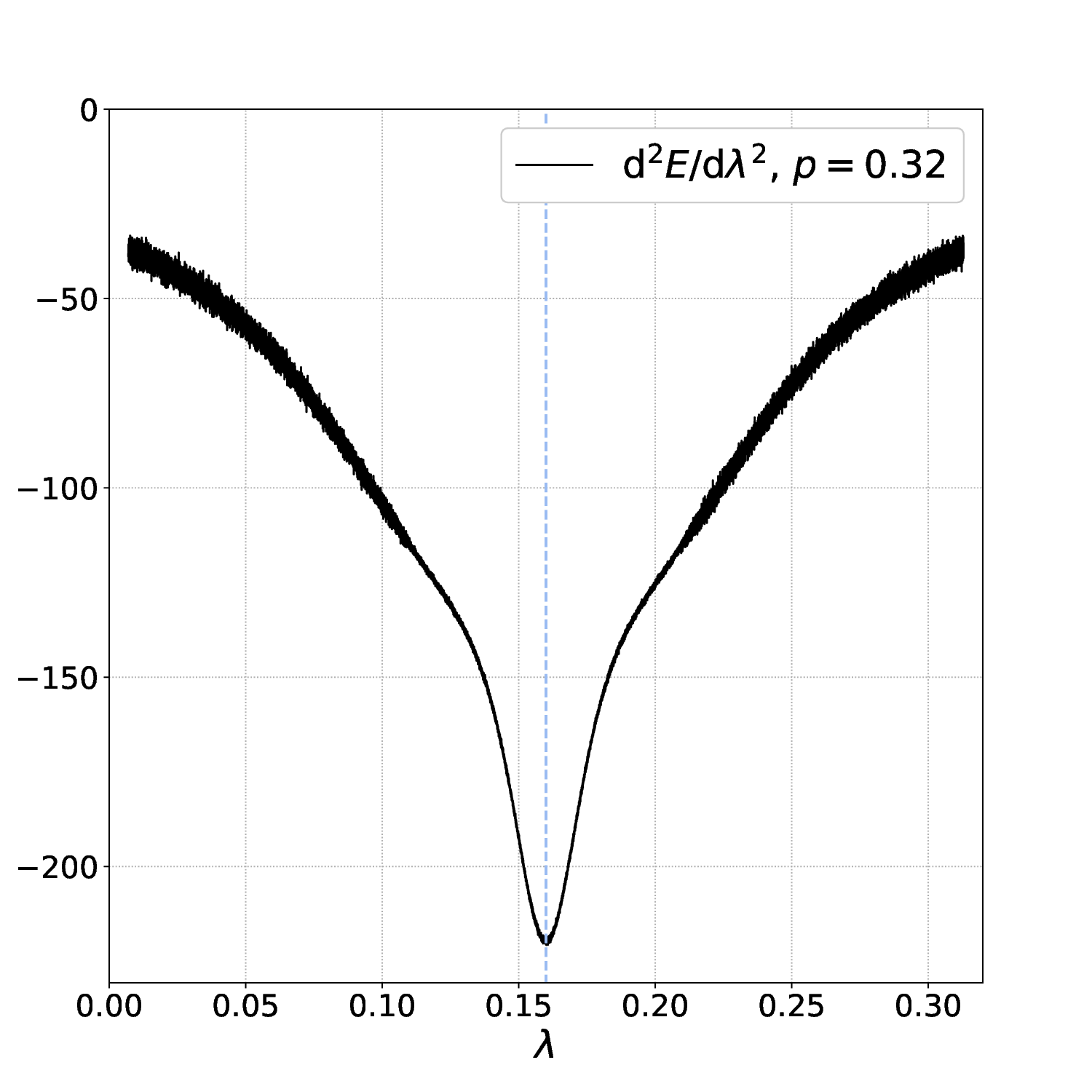}
}
 {
\includegraphics[width=6cm]{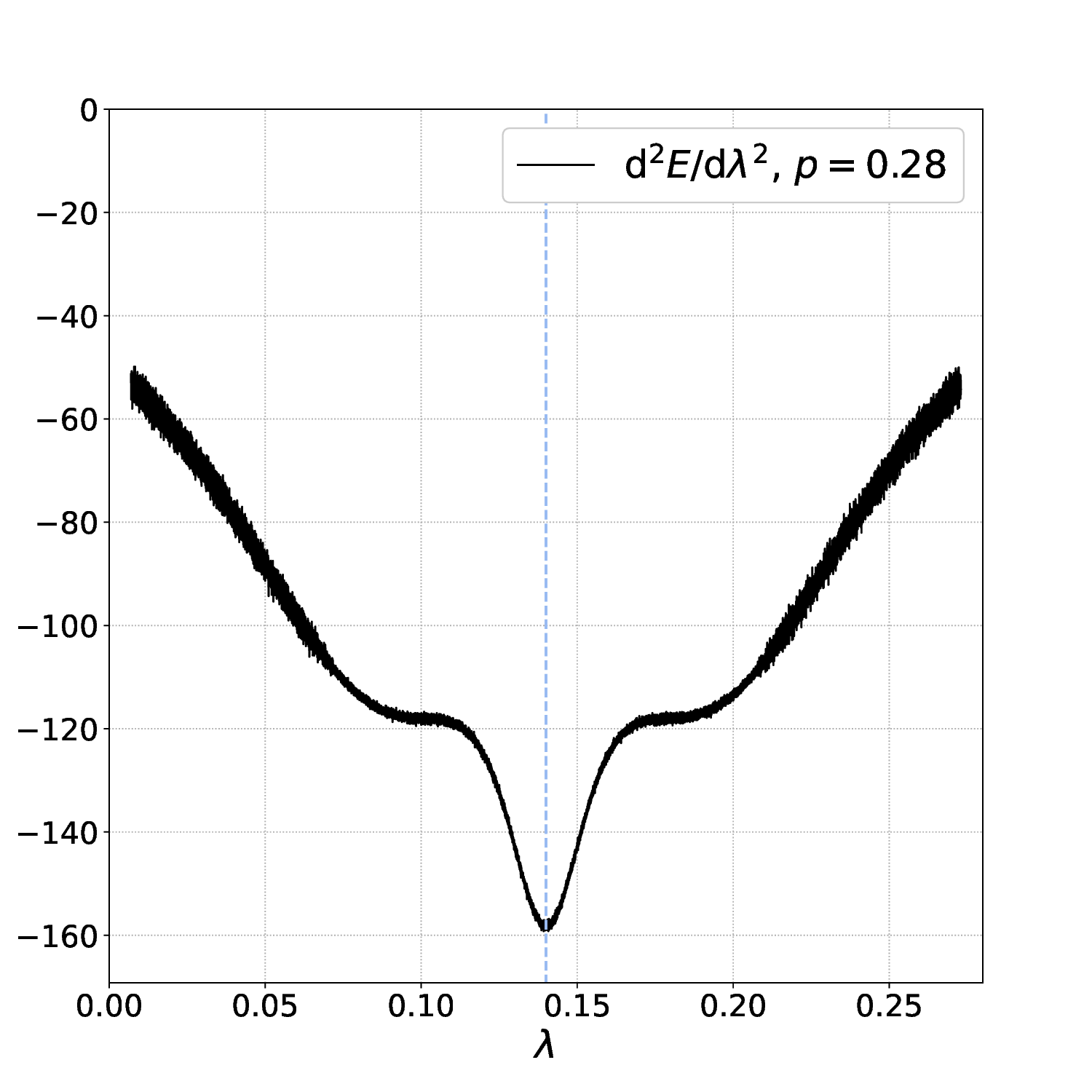}
}
 {
\includegraphics[width=6cm]{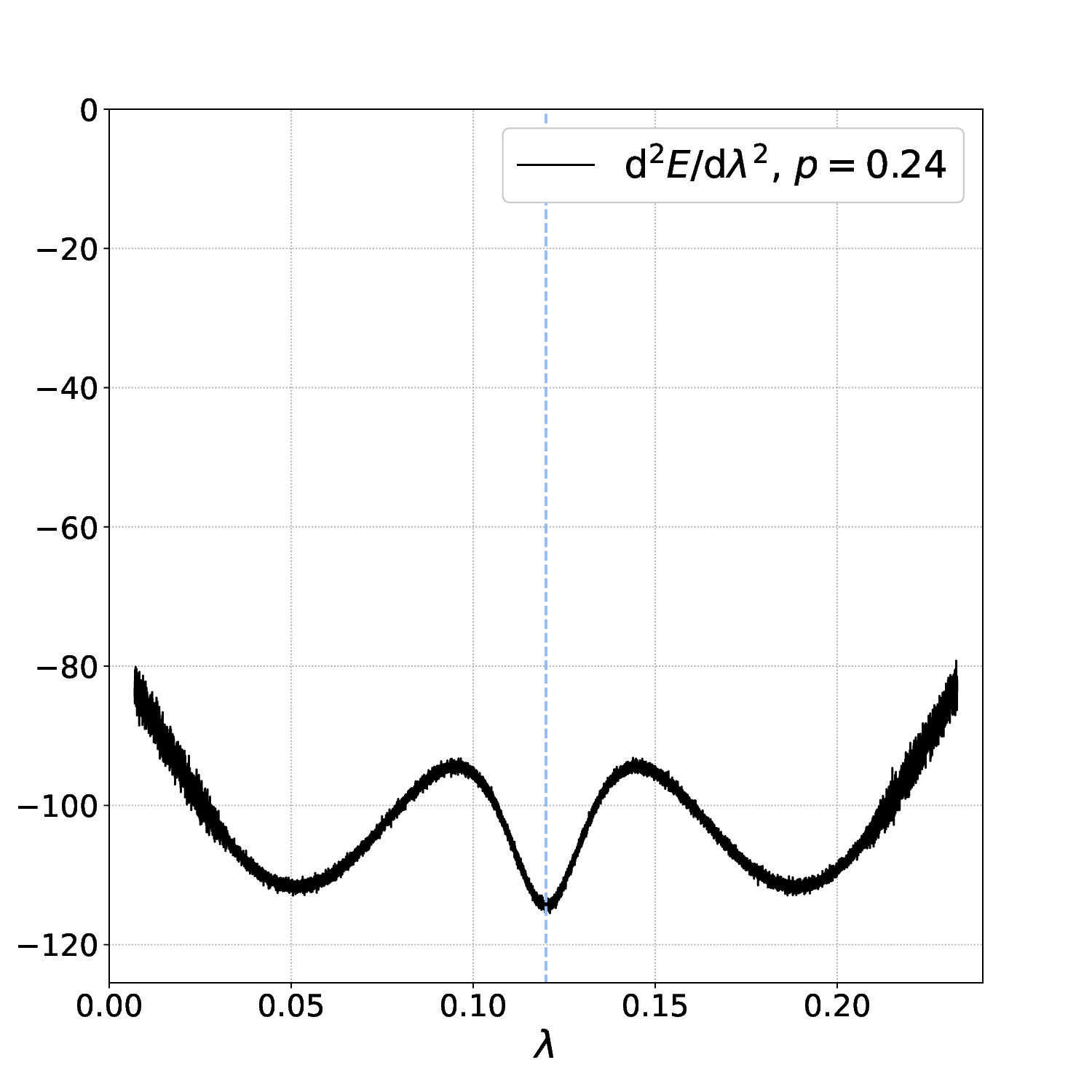}
}
\caption{\label{3maxima} Energy as a function of $\lambda$, for blue paths with different values of $p$: (a) $p=0.36$,
(b) $p=0.32$, (c) $p=0.28$, and (d) $p=0.24$.}
\end{figure}

Besides, no matter what value $p$ we choose for the path  of the solid blue line, we can always observe
an extremal point of $\mathrm{d}^2E/dg^2$ right on the dual line, as shown in  Fig.~\ref{3maxima}. However, an extremal point does not
have to be a critical point. To determine the end of the first-order phase transition line outside the deconfined phase, we need to investigate the DOS of $\tilde{Z}$.

Since the Higgs phase and confined phase are continuously connected when
$g$, $\lambda\to\infty$ \cite{PhysRevD.19.3682}, the first-order line should vanish somewhere. Figure \ref{dos_dual} shows the variation of the DOS on the dual line and gives two special points,
$g_1=0.30$ and $g_2=0.65$. We consider $g_1$ as the end of the line because the maximum  at
$\tilde{Z}=-10$ suggests that the system starts to mainly permit the excitation of the smallest loop, which is composed of four spins, while other possible configurations are generally disfavored. Consequently, the fusion of the flux and charge, which is a fermion in this model, vanishes and the system turns into confinement \cite{PhysRevB.67.245316}.

  \begin{figure}[htb] 
 \center{\includegraphics[width=8cm]  {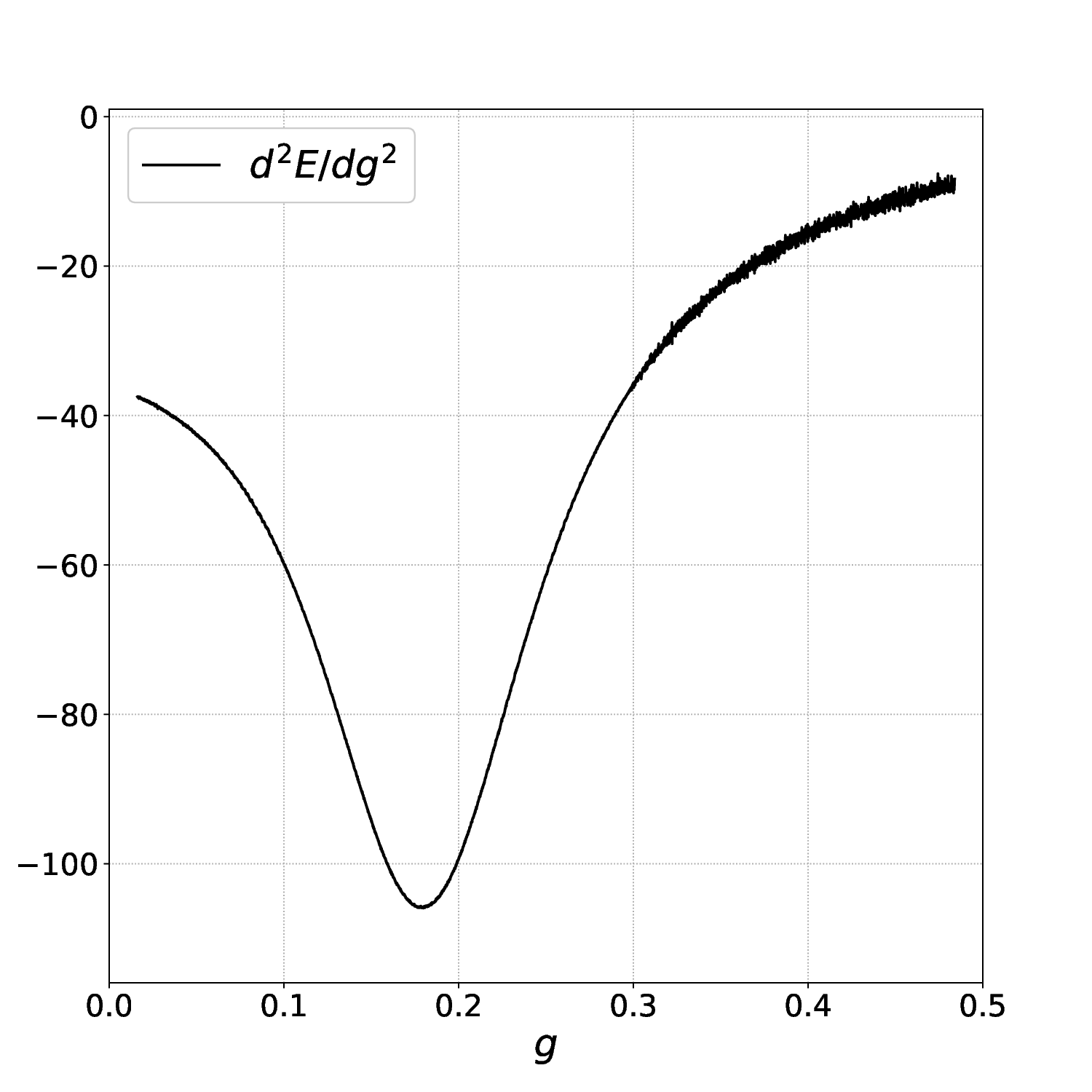}}
 \caption{\label{dual_evo}The second derivative of energy on the dual line, with  an extremal point at $g=0.1798<0.24$.}
 \end{figure}

Besides this, Fig.\ \ref{phasediagram} suggests that  there seem to be  three extremal points on the paths in solid blue lines when $0.19<p<0.48$. However, this feature  can only be clearly observed when $p<0.28$. For $p>0.28$, we observe a single extremal point on the dual line. We regard this as a finite-size effect, which makes the extremal point on the dual line indistinguishable from  the two on the second-order lines.

The extremal points on the dual line make it natural for us to study the energy   right on the dual line, so we choose a new path from $(0,0.5)$, through $(0.5,0.5)$, and to $O(0,0)$. Surprisingly we find an extremal point at $g_3=0.1798$, which is in
the deconfined phase (See Fig. \ref{dual_evo}). Moreover,  the DOS is also disordered on the part of the dual line continued from the first-order line into the deconfined phase; as indicated in Fig.~\ref{dos_sec},  it is likely  that the first-order line   extends for a range in the deconfined phase and ends at $g_3$ (See Fig. \ref{phasediagram}).

The phase diagram we obtain is similar to the diagram in Ref. \cite{Vidal:2008uy}, which is derived by using perturbation theory. The difference lies in the extension of the first-order line in the deconfined phase. The region near the tricritical point has not been clear. Our result corresponds to  one of the possibilities proposed in Ref. \cite{PhysRevB.82.085114}.

\section{Error analysis}
\label{sec:trotter}

As the Trotter-Suzuki decomposition is used in each step, the upper bound of
the total error is a summation of the errors in each step.

We take the subpath from $R(p,r)=(0.5,0.05)$ to $R'(0,r)=(0,0.05)$ as an example. As derived in the Appendix, the accumulated Trotter error $\epsilon$   is
\begin{equation}
\begin{split}
\epsilon  \lesssim
&{(\Delta t)^3}N  (13.5p+54p r^2+54pr+54r^2 \\
&+9p^2+18p^2r+13.5r), \\
\end{split}
\end{equation}
where $\Delta t$ is the time for each step of varying the parameter values, and $N$ is the number of steps.

For variation of $g$ from $0$, $\Delta g$ is the variation in each step; $\Delta g \propto \Delta t$. We choose $\Delta g=10^{-5}$ and $\Delta t=2\times 10^{-2}$,  giving $\epsilon \approx 4.581$, which seems to be too large for us. However, this error bound is obtained by adding  the absolute values of the errors in all steps, and the actual  errors are not  always positive and   may cancel  each other \cite{PRXQuantum.2.010323}. We  demonstrate this  in  the following steps.

\begin{figure}[h] 
 \center{\includegraphics[width=8cm]  {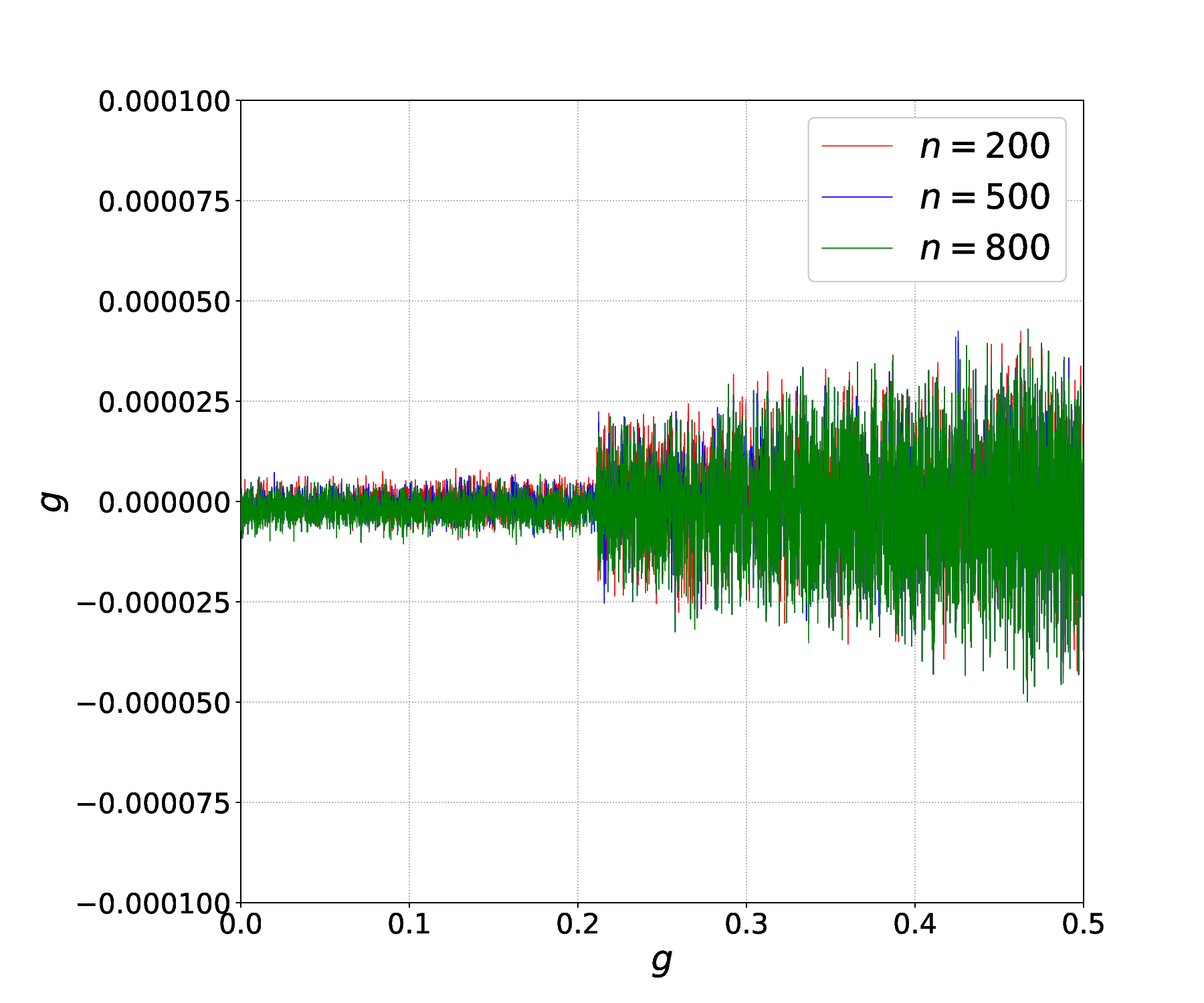}}
 \caption{\label{error}The diagram of the difference $E_{n_0}(g) -  E_n(g)$  for  different values of $n$.}
 \end{figure}

Rewrite $\Delta g = 0.0001/n$ and $ \Delta t=0.2/n$, then $n=n_0=10$ under the parameter values above. Define
\begin{equation}
\begin{split}
&\delta_{n} = \sum_{g=0}^{p} \vert E_{n_0}(g) -  E_n(g) \vert, \\
&\alpha_{n} = \max_{g\in[0,p]} \vert E_{n_0}(g) -  E_n(g) \vert. \\
\end{split}
\end{equation}
where $\delta_n$ is a 1-norm of the difference between two energy functions $E_{n_0}(g)$ and
 $E_n(g)$, and $\alpha_n$ represents the largest value of the difference. The results in Tabel \ref{contrast_tab} and Fig. \ref{error} show that even when  $\varepsilon$ at $n_0=10$ is around 6363 times more than that at $n=800$, $\delta_{n}$ is merely $0.04075$, and the maximal  deviation $\alpha_n$  is only $0.005\%$. This is  direct evidence that the error at each step cannot always be positive; thus, our choice of $n=10$ is reasonable.

\begin{table}[h]  \centering
\begin{tabular*}{8.5cm}{cccccc}
\hline
$n$ & $\Delta g$  & $\Delta t$ & $\varepsilon$  & $\delta_{n}$	&$\alpha_{n}$\\
\hline
10 & $1.00\times 10^{-5}$ & $2.0\times 10^{-2}$ & 4.58100 & 0 & 0\\
200  & $5.00\times 10^{-7}$  & $1.0\times 10^{-3}$& 0.01145  & 0.03982 & $ 4.79\times 10^{-5}$\\
500  & $2.00\times 10^{-7}$ & $4.0\times 10^{-4}$ & 0.00183 & 0.04009 & $ 4.99\times 10^{-5}$\\
800  & $1.25\times10^{-7}$ & $2.5\times10^{-4}$ & 0.00072 & 0.04075 & $4.98\times 10^{-5}$\\
\hline
\end{tabular*}
 \caption{\label{contrast_tab}Different values of $\delta_{n}$ and $\alpha_{n}$ for $n=10,200,500,800$.}
\end{table}

The results also demonstrate that the Trotter steps can actually be reduced in an actual quantum simulation.

 \section{Comparison  with  Exact Diagonalization }\label{sec:cmp}
To verify the validity of our approach, we compare the results on
a $2\times 2$ lattice on a torus in both exact diagonalization  and our approach. We  do not directly make  exact diagonalization   on a $3\times 3$ lattice, as it would need much more computing resources, while  the scale of
$2\times 2$ is close to $3\times 3$.

For this model, exact diagonalization is more time-consuming than our approach, in which  one does parallel computing with GPU. Specifically, for red paths
with $\Delta g=10^{-5}$, the  time in our approach  is within 30 minutes on a single GeForce RTX 3090 for 19 qubits (complex data composed of a double data type for real components and imaginary components).

\begin{figure}[htb] 
\centering
{
\includegraphics[width=6cm]{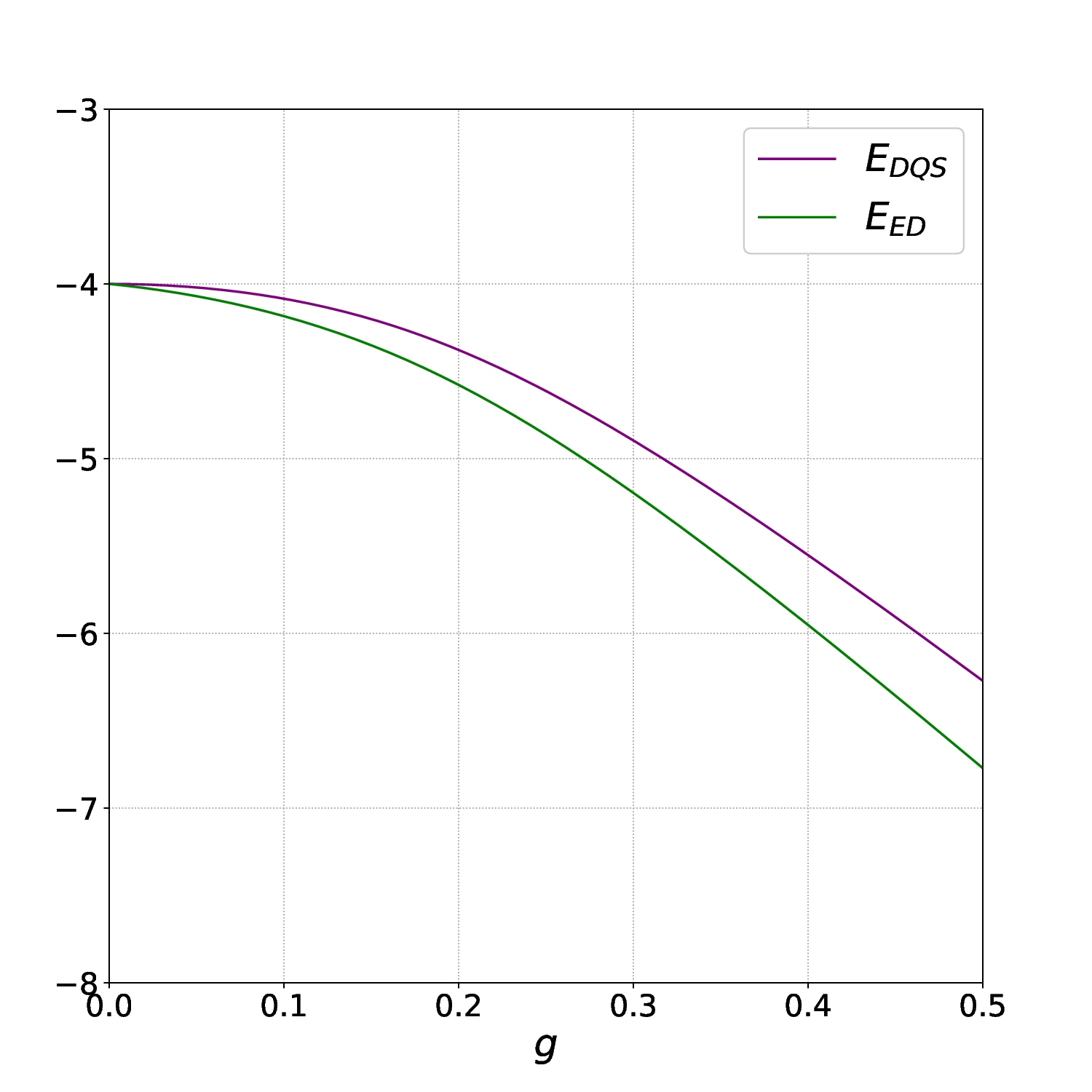}
}
 {
\includegraphics[width=6cm]{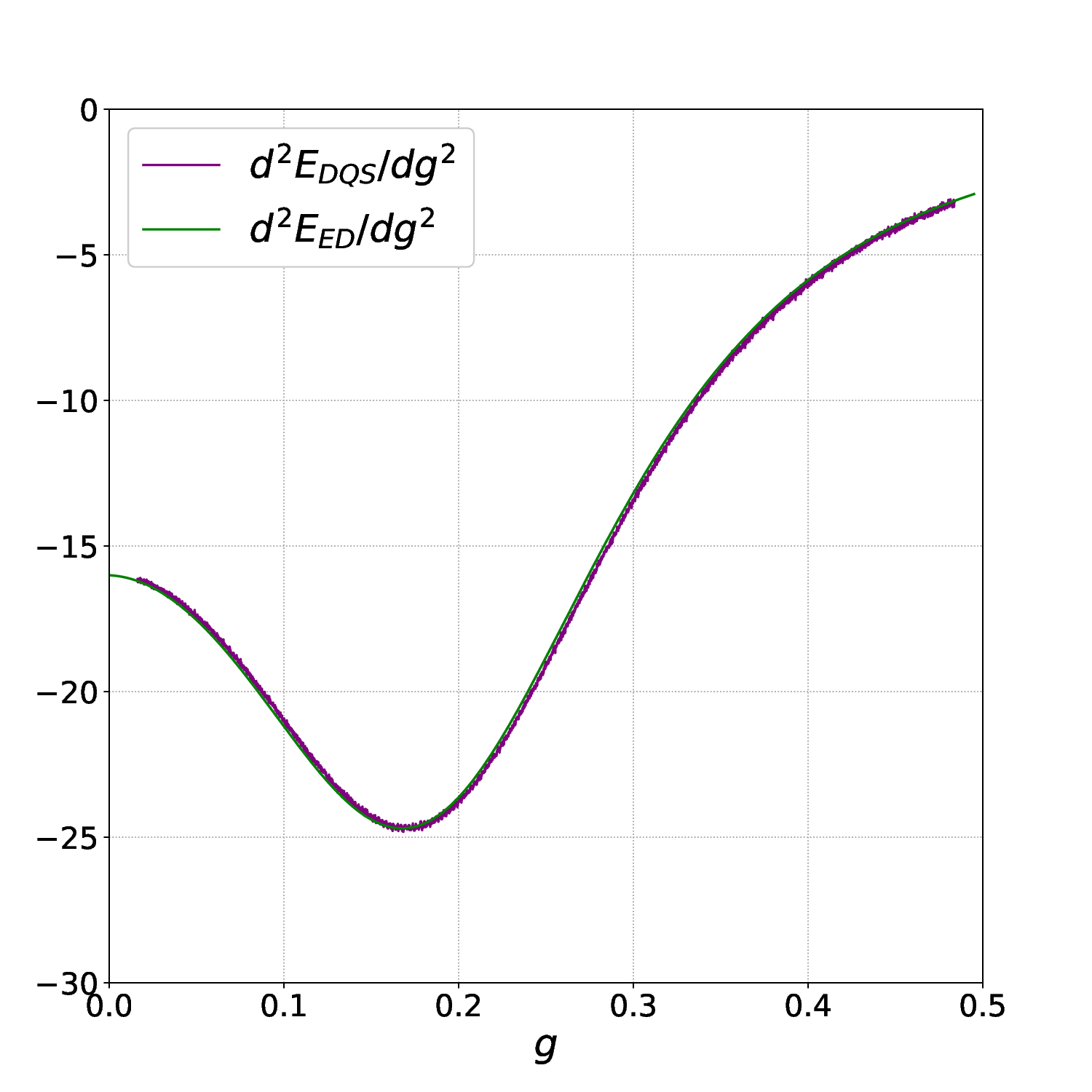}
}
    \caption{\label{2by2_OP}The energy  and its second-order
    derivative from $(0,0)$ to $(0.5, 0)$ by using ED and  pseudoquantum simulation, respectively.}
\end{figure}

We use the same parameter values as  in our $3\times 3$ model. For the evolution from $O(0, 0)$ to $P(0.5, 0)$, the comparison is shown in
Fig. \ref{2by2_OP}. The unavoidable Trotter error and nonadiabatic error in DOS result in a difference between $E_{\mathrm{ED}}$ and $E_{\mathrm{DQS}}$, the values obtained in exact diagonalization and in our classical demonstration of digital quantum simulation, dubbed pseudoquantum simulation, respectively.  However, the results on the  second derivatives in the two approaches are very close to each other. As the quantum phase transition behavior is largely investigated through the second derivatives of energy, the comparison confirms the reliability of our approach.

\begin{figure}[htb] 
   \centering
{
\includegraphics[width=6cm]{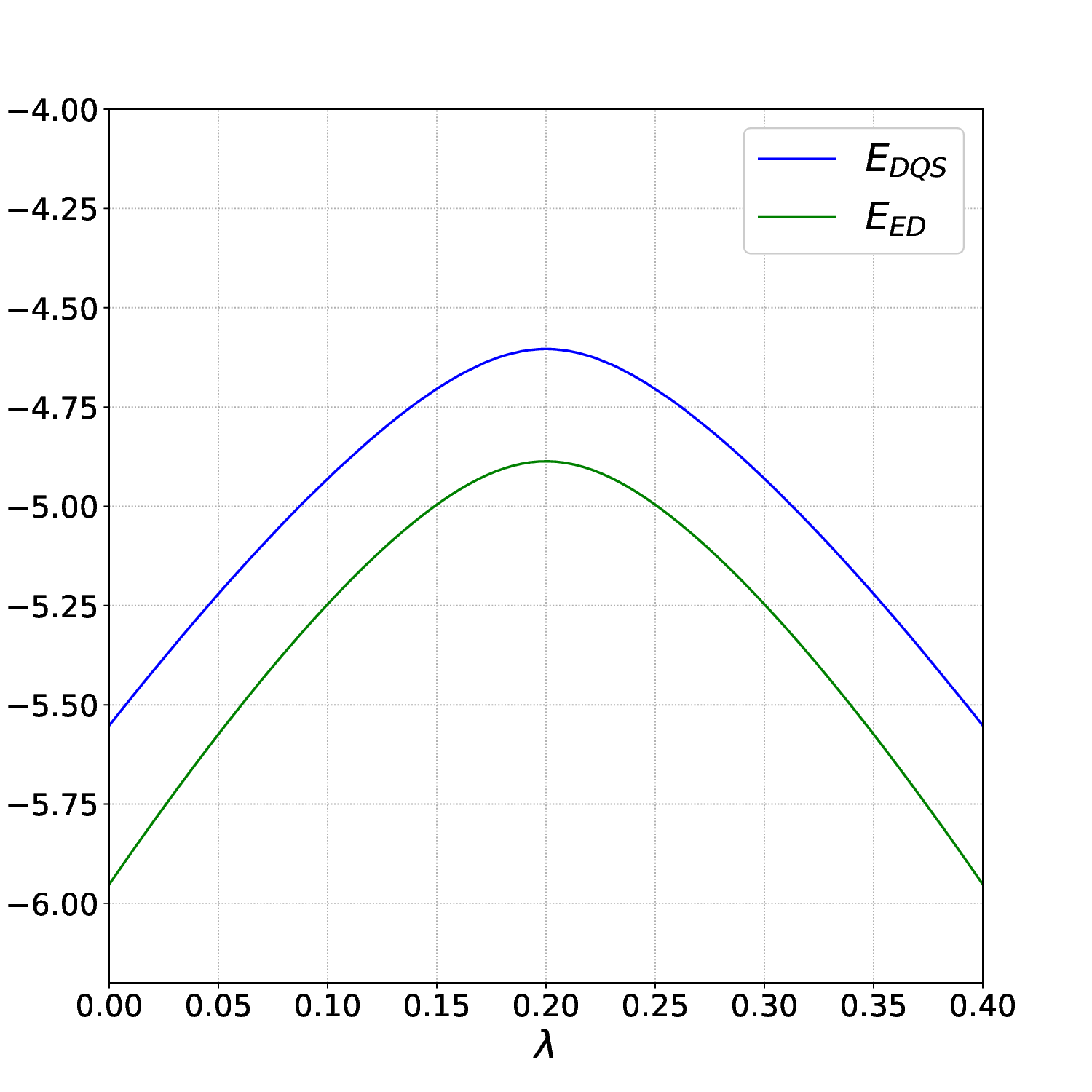}
}
 {
\includegraphics[width=6cm]{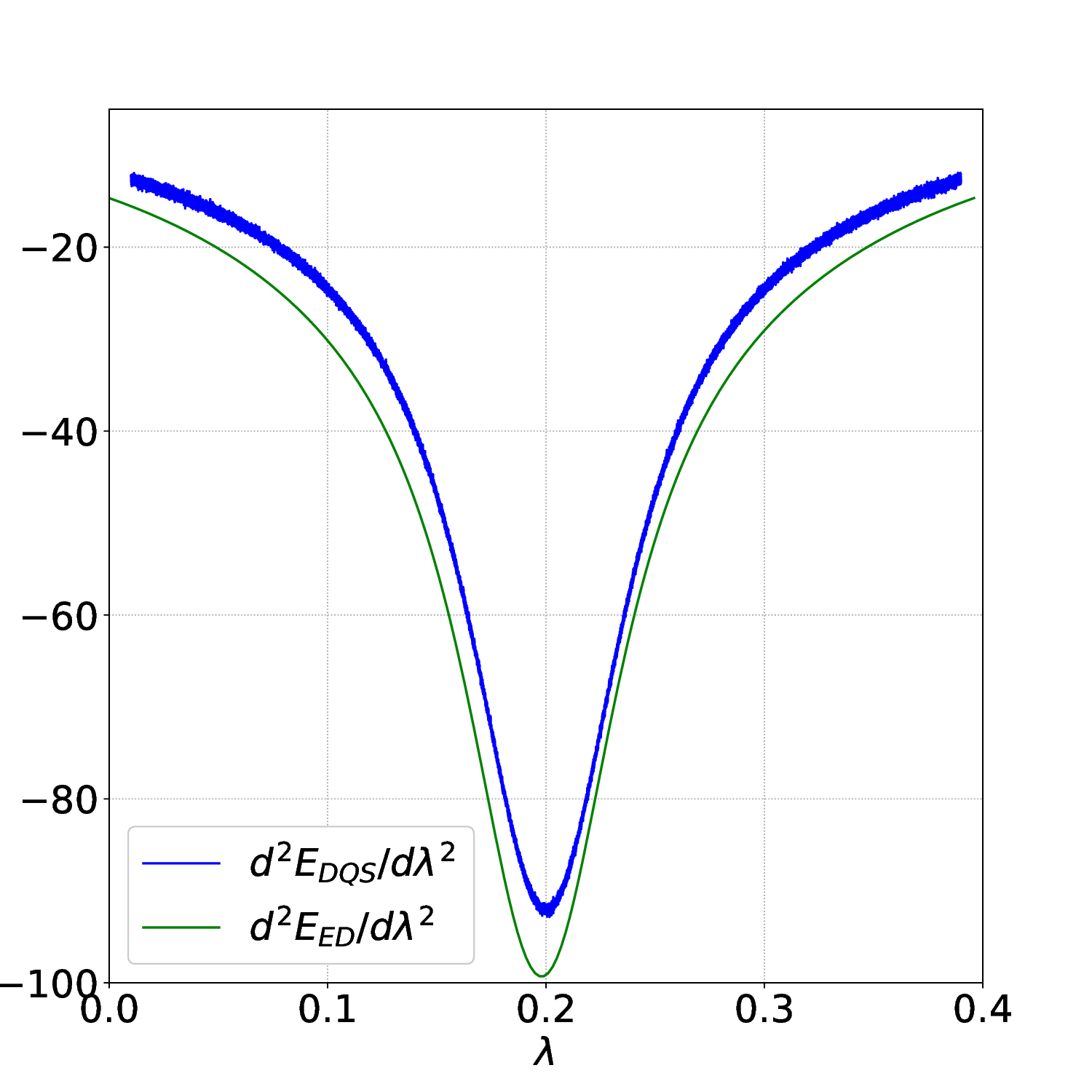}
}
    \caption{\label{blue_example}The energy  and its second-order
    derivative from $(0.4,0)$ to $(0, 0.4)$ by using ED and   pseudoquantum simulation.}
\end{figure}

As shown  in Fig. \ref{blue_example}, similar results can also be observed  outside the
deconfined phase—for example, along a blue path in blue, with $p=0.4$.

\begin{figure}[htb] 
       \centering
{
\includegraphics[width=6cm]{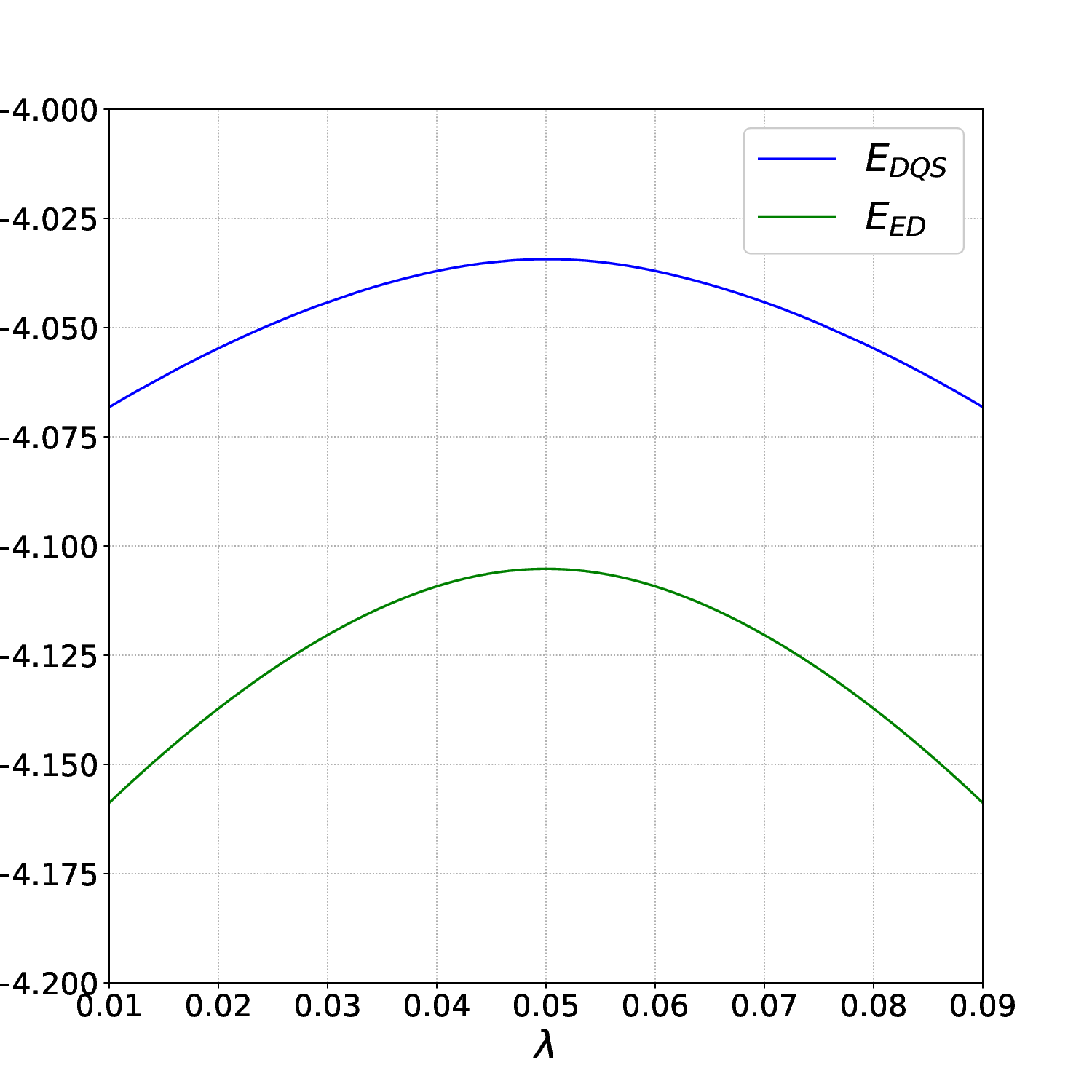}
}
 {
\includegraphics[width=6cm]{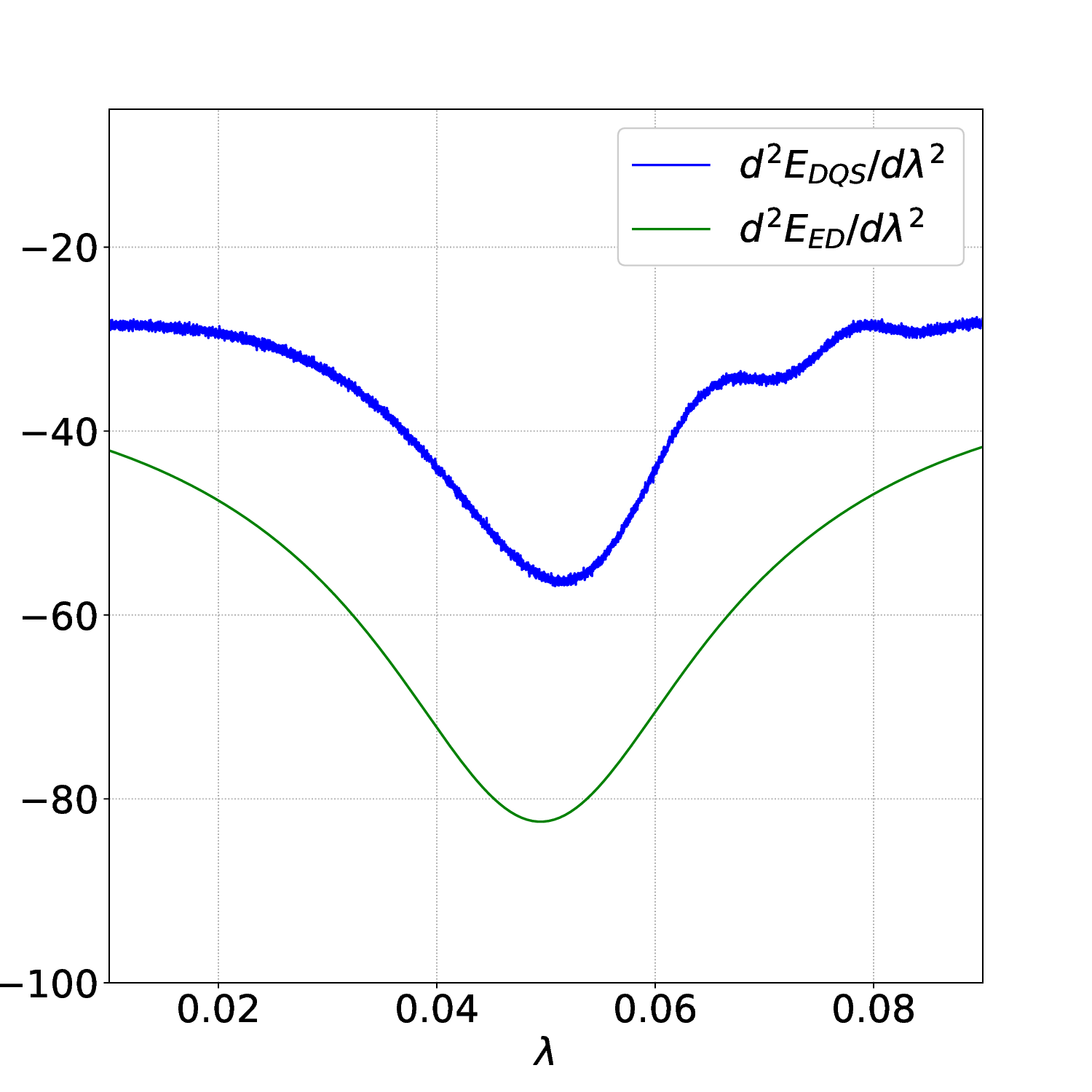}
}
    \caption{\label{blue_example2}The energy curve and its second-order
    derivative from $(0.1,0)$ to $(0, 0.1)$ by using ED and   pseudoquantum simulation.}
\end{figure}

If the evolution path passes the deconfined phase, which is topological,  on a  blue path with $p=0.1$,  for
example, the shapes of $E_{\mathrm{ED}}$ and $E_{\mathrm{DQS}}$
are quite different, as shown in  Fig. \ref{blue_example2}).

For a  smaller $p$,
even the shape of ${E_{\mathrm{DQS}}}$ is  clearly asymmetric, which means that
the difference of $E_{\mathrm{DQS}}$ on $P$ and $P'$ is fairly large.  The asymmetry cannot be observed in exact diagonalization  or  QMC.

On a
$2\times 2$ lattice, as in the case of  a $3\times 3$ lattice, the $E_{DQS}$ curves  on the path from  $P'$ to $P$ cross that   on the reverse path from  $P'$ to $P$.

\section{Summary}
\label{sec:sum}

We have  designed the quantum simulation scheme of the $\mathbb{Z}_2$ gauge-Higgs model. The quantum simulation scheme is digital, as it is based on Trotter decomposition of the unitary evolution.  It is also based on a quantum adiabatic algorithm. For each  parameter value, which is varied slowly,  the Trotter decomposition is used in executing the unitary transformation. Within each step in the Trotter decomposition, the unitary transformations are realized in terms of simple quantum circuits.

Moreover, as the quantum computers nowadays have not been capable of such quantum simulations, we make a thorough classical demonstration by  using   QuEST simulator  on an NVIDIA GeForce RTX 3090 GPU server. This so-called pseudoquantum simulation  not only facilitates the development of algorithms for future real quantum simulation, but is also  a numerical method.

Then, we make a thorough classical demonstration  by  using  the QuEST simulator  in a NVIDIA GeForce RTX 3090 GPU server.  Our demonstration is on  a $3\times 3$ lattice,   as limited by computational time in the pseudoquantum simulation and the number of qubits that can be realized in present quantum hardware. However, we have verified the reliability of our approach by comparing with the exact diagonalization on a $2\times 2$ lattice.

We have obtained some clear results on the
topological properties of the deconfined phase, which appears useful for the solution of some open questions. In particular, our work suggests  that the two lines of second-order transitions meet on the line of the first-order transition, but not on its end.

\begin{acknowledgments}
This work was supported by the National Natural Science Foundation of China (Grant No. 12075059).
\end{acknowledgments}

\appendix
\section*{APPENDIX: Calculation of the  Trotter errors}

\label{calc}

It is useful to calculate $\lVert H_j \rVert$ and $\lVert [H_k, H_l] \rVert$ in advance where $\lVert \cdot \rVert$ denotes the spectral norm of an operator, or the largest singular value of it. We use the estimation that $\lVert O \rVert \sim 1$, where $O$ is a
 Pauli operator or a tensor product of Pauli operators.

\begin{figure}
 \center{\includegraphics[width=6cm]  {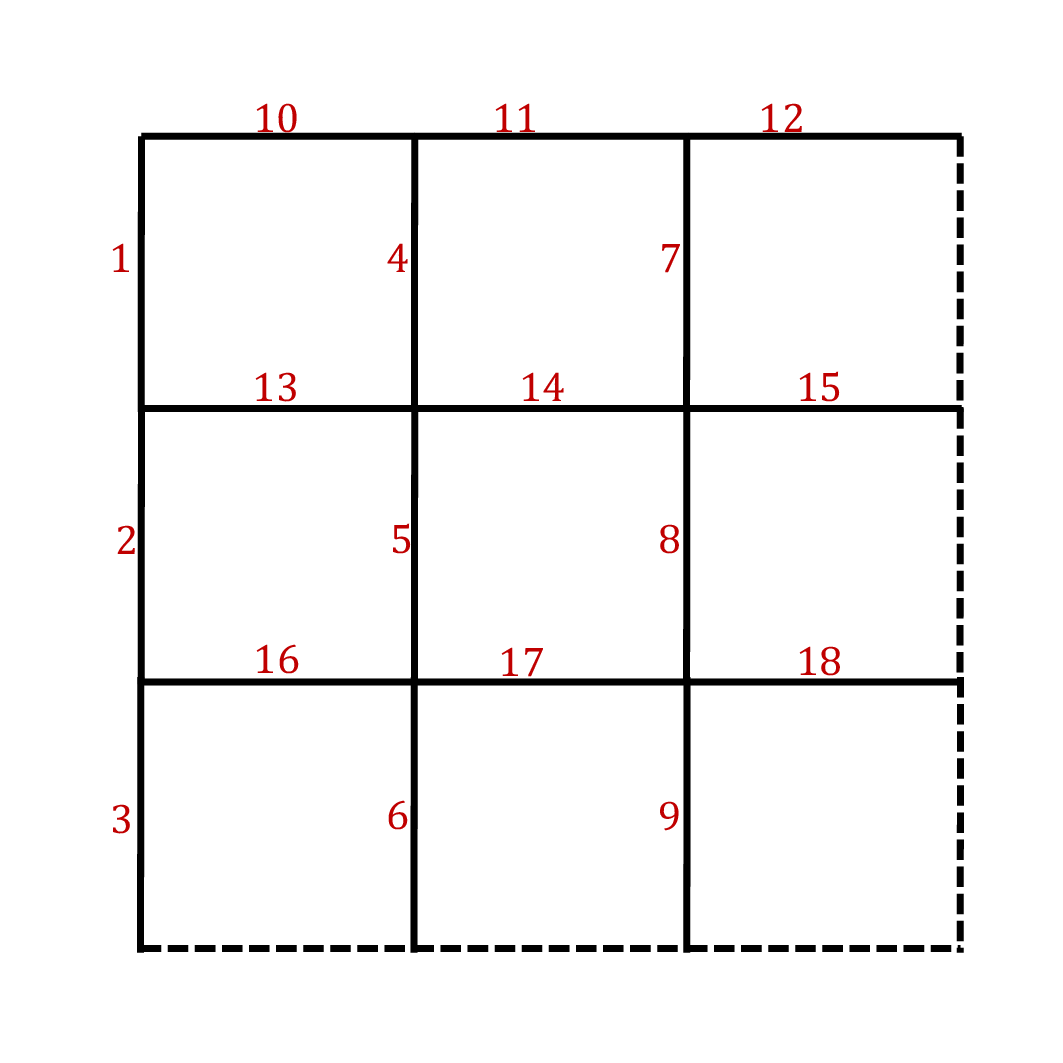}}
 \caption{\label{lattice}Lattice  used in the simulations.}
 \end{figure}

For our specific $3\times3$ lattice model (see Fig. \ref{lattice}), we have
\begin{equation}\label{EachTerm}
\begin{split}
&\lVert H_1 \rVert \sim 18g, \quad \lVert H_3 \rVert\sim18\lambda,\\
&\lVert H_2 \rVert \leq \frac{1}{2} \lVert \sum_p B_p^z \rVert  + \frac{1}{2} \lVert \sum_v A^x_v\rVert   \sim9,
\end{split}
\end{equation}
and
\begin{equation}\label{cmt12}
\begin{split}
&\lVert[H_1,H_2]\rVert\\
 =&\lVert[-g\sum_l\sigma^x_l,-\frac{1}{2}\sum_pB_p^z-\frac{1}{2}\sum_vA_v^x]\rVert\\
 =&\frac{g}{2}\lVert[\sum_l\sigma^x_l,\sum_pB_p^z]\rVert\\
 \leq&\frac{g}{2}\sum_l\lVert[\sigma^x_l,\sum_pB_p^z]\rVert\\
 =&18\frac{g}{2}\lVert[\sigma^x_4,\sum_pB_p^z]\rVert\\
  =&18\frac{g}{2}\lVert[\sigma^x_4, \sigma^z_1\sigma^z_{10}\sigma^z_4\sigma^z_{13}]+[\sigma^x_4, \sigma^z_4\sigma^z_{11}\sigma^z_7\sigma^z_{14}]\rVert\\
 \sim&18{g},
\end{split}
\end{equation}
where the coefficient $18$ comes up due to the symmetry of the lattice. Similarly, we have $\lVert[H_2,H_3]\rVert\sim 18\lambda$ and
\begin{equation}\label{cmt13}
\begin{split}
&\lVert[H_1,H_3]\rVert\\
 =&\lVert[-g\sum_l\sigma^x_l,-\lambda\sum_s\sigma_s^z]\rVert\\
 =&g\lambda\lVert[\sum_l\sigma^x_l,\sum_s\sigma_s^z]\rVert\\
 =&g\lambda\lVert\sum_l[\sigma_l^x,\sigma^z_l]\rVert\\
  \sim&36g\lambda.
\end{split}
\end{equation}

Multiplying a minus sign does not change the eigenvalues of a matrix; thus, $\lVert [H_2,H_1]\rVert\sim18g$, $\lVert [H_3,H_1]\rVert\sim36g\lambda$ and $\lVert [H_3,H_2]\rVert\sim18\lambda$.

 Suppose  $H=\sum_{\gamma=1}^{\Gamma} H_{\gamma}$; then, the tight error bound for the second-order decomposition is \cite{PhysRevX.11.011020}
\begin{equation}
\begin{split}
&\lVert S_2(\Delta t)-e^{-i(\Delta t)H} \rVert\\
 \leq& \frac{(\Delta t)^3}{12}\sum_{\gamma_1=1}^{\Gamma}  \lVert [ \sum_{\gamma_3=\gamma_1+1}^{\Gamma} H_{\gamma_3},[\sum_{\gamma_2=\gamma_1+1}^{\Gamma}H_{\gamma_2},H_{\gamma_1}]]\rVert\\
&+\frac{(\Delta t)^3}{24}\sum_{\gamma_1=1}^{\Gamma}\lVert [H_{\gamma_1}, [H_{\gamma_1}, \sum_{\gamma_2=\gamma_1+1}^{\Gamma}H_{\gamma_2}]].
\rVert\\
\end{split}
\end{equation}

Our Hamiltonian is $H=H_1+H_2+H_3$; thus, the error bound  is
\begin{equation}\label{T2S3}
\begin{split}
&\lVert S_2(\Delta t)-e^{-itH} \rVert \\
\leq& \frac{(\Delta t)^3}{12}\sum_{l=1}^{3}\lVert [ \sum_{n=l+1}^{3} H_n,[\sum_{m=l+1}^{3}H_m,H_l]]\rVert\\
&+\frac{(\Delta t)^3}{24}\sum_{l=1}^{3}\lVert [H_l, [H_l, \sum_{m=l+1}^{3}H_m]]  \rVert\\
\leq&\frac{(\Delta t)^3}{12}\sum_{l=1}^3\sum_{n=l+1}^{3} \sum_{m=l+1}^{3} \lVert  H_n[H_m,H_l]- [H_m,H_l]H_n   \rVert\\
&+\frac{(\Delta t)^3}{24}\sum_{l=1}^{3} \sum_{m=l+1}^{3}  \lVert H_l [H_l, H_m] -[H_l, H_m]H_l\rVert\\
\leq&\frac{(\Delta t)^3}{6}\sum_{l=1}^3\sum_{n=l+1}^{3} \sum_{m=l+1}^{3}    \lVert H_n\rVert \lVert [H_m,H_l] \rVert\\
&+\frac{(\Delta t)^3}{12}\sum_{l=1}^{3} \sum_{m=l+1}^{3}  \lVert H_l \rVert   \lVert [H_l, H_m\rVert\\
=&\frac{(\Delta t)^3}{6}W_1+\frac{(\Delta t)^3}{12}W_2,
\end{split}
\end{equation}

where

\begin{equation}
\begin{split}
&W_1=\sum_{l=1}^3\sum_{n=l+1}^{3} \sum_{m=l+1}^{3}    \lVert H_n\rVert \lVert [H_m,H_l] \rVert,\\
&W_2=\sum_{l=1}^{3} \sum_{m=l+1}^{3}  \lVert H_l \rVert   \lVert [H_l, H_m]\rVert.\\
\end{split}
\end{equation}

According to Eq. (\ref{EachTerm}), we have
\begin{equation}\label{W1}
\begin{split}
W_1=&\sum_{n=2}^{3} \sum_{m=2}^{3}    \lVert H_n\rVert \lVert [H_m,H_1] \rVert +
\lVert H_3\rVert \lVert [H_3,H_2] \rVert\\
=& \lVert H_2\rVert( \lVert [H_2,H_1] \rVert +\rVert \lVert [H_3,H_1] \rVert)\\
&+  \lVert H_3\rVert( \lVert [H_2,H_1] \rVert
+ \lVert [H_3,H_1] \rVert
+\lVert [H_3,H_2] \rVert)\\
\sim&9(18{g}+36g\lambda)
+18\lambda(18{g}
+36g\lambda + 18{\lambda})\\
=&162({g}+2g\lambda)
+324\lambda({g}+2g\lambda+ {\lambda})\\
=&162g+648g\lambda^2+648g\lambda+324\lambda^2,
\end{split}
\end{equation}

and
\begin{equation}\label{W2}
\begin{split}
W_2=&\sum_{m=2}^{3}  \lVert H_1 \rVert   \lVert [H_1, H_m\rVert +  \lVert H_2 \rVert   \lVert [H_2, H_3\rVert\\
=&  \lVert H_1 \rVert  ( \lVert [H_1, H_2\rVert + \lVert [H_1, H_3\rVert) +  \lVert H_2 \rVert   \lVert [H_2, H_3\rVert\\
\sim&18g(18g + 36g\lambda)+9(18{\lambda})\\
=&324g^2+648g^2\lambda + 162\lambda.
\end{split}
\end{equation}

The evolution in Sec. \ref{sec:trotter} is from $(0,r)$ to $(p,r)$ on $l:\lambda=r$ with $g$ increasing from $0$ to $p$ in $N = p/\Delta g$ steps. Denote $g=n\Delta g$; then the total error is
\begin{equation}  \label{target}
\begin{split}
 \epsilon  \leq
 &\frac{(\Delta t)^3}{6} \sum^{N}_{n=1} W_{1,n} + \frac{(\Delta t)^3}{12}\sum_{m=1}^{N}\Delta_mW_{2,m}\\
\sim  &\frac{(\Delta t)^3}{6} \sum^{N}_{n=1}(162g+648g\lambda^2+648g\lambda+324\lambda^2) \\
&+ \frac{(\Delta t)^3}{12}\sum_{m=1}^{N} (324g^2+648g^2\lambda + 162\lambda)   \\
=&{(\Delta t)^3} \sum^{N}_{n=1}(27g+108g\lambda^2+108g\lambda+54\lambda^2) \\
&+ {(\Delta t)^3}\sum_{m=1}^{N} (27g^2+54g^2\lambda+13.5\lambda)   \\
=&{(\Delta t)^3} \sum^{N}_{n=1}(27g+108g\lambda^2+108g\lambda+54\lambda^2 \\
&+27g^2+54g^2\lambda+13.5\lambda)  \\
=&{(\Delta t)^3} \sum^{N}_{n=1}(27g+108gr^2+108gr+54r^2 \\
&+27g^2+54g^2r+13.5r).   \\
 \end{split}
\end{equation}

Besides this, we have
\begin{equation}
\begin{split} \label{g^1}
&\sum_{n=1}^{N} g = \sum_{n=1}^{N} (n\Delta g)
\approx \Delta g \frac{N^2}{2} =  \frac{1}{2} pN,   \\
&\sum_{n=1}^{N} g^2 = \sum_{n=1}^{N} ( n\Delta g)^2
\approx \Delta g^2 \frac{N^3}{3}
= \frac{1}{3} p^2N.  \\
 \end{split}
\end{equation}

Rewriting Eq. (\ref{target}) with Eq. (\ref{g^1}), we have
\begin{equation}
\begin{split}
\epsilon \lesssim
&{(\Delta t)^3}  (27\frac{1}{2}pN+108\frac{1}{2}pN r^2+108\frac{1}{2}pNr  \\
&+54r^2N +54\frac{1}{3}p^2N+54\frac{1}{3}p^2Nr+13.5rN)  \\
=&{(\Delta t)^3}N  (13.5p+54p r^2+54pr+54r^2 \\
&+9p^2+18p^2r+13.5r).  \\
\end{split}
\end{equation}





\end{document}